VIRGINIA POLYTECHNIC INSTITUTE AND STATE UNIVERSITY

# The Agile Adoption Process Framework

## *Detailed Reference Document*

Agile Practices and Concepts with Assessment Indicators

*including a special Appendix about the 5 Agile Levels*

*By: Ahmed Sidky*

# Table of Contents

# Overview of the Process Framework

*Pictorial Representation*

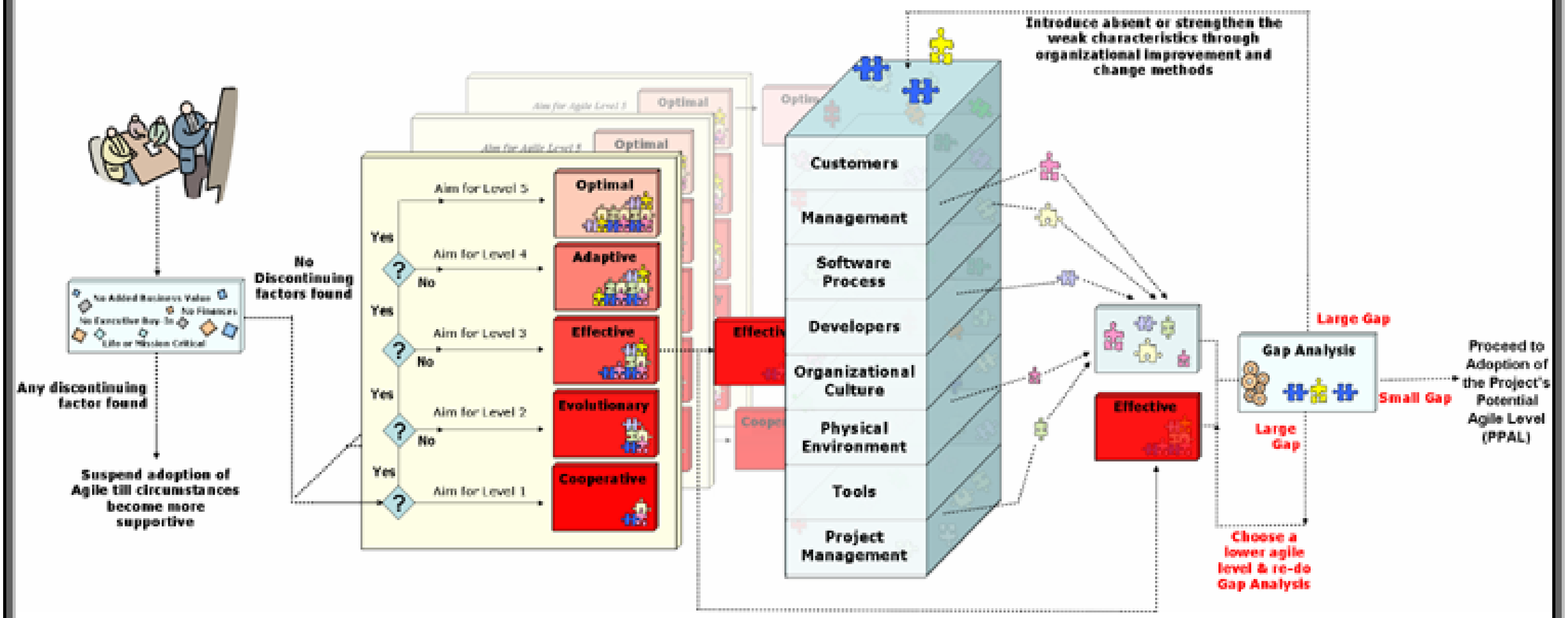

Stage 1 — Stage 2 — Stage 3

**Figure 1. Overview of the Process Framework**

# *Overall Description*

The objective of this research is to provide organizations with a process that can guide projects in adopting more agile software development approaches. The result of this research is a process framework consisting of various components that work together to provide a roadmap and guide for organizations wishing to adopt agile practices during project development. Figure 1 illustrates the stages and sequence of events in this process framework.

The objective of the First stage of the process framework is to identify any organizational factor that would prevent the adoption of agile practices. The stage starts with an introductory meeting with the key stakeholders of the Agile adoption endeavor. This meeting focuses on explaining the main concepts behind Agile methods and presents an overview of the stages that the adoption effort will go through. After this informational session the agile consultant examines the organization for the presence of any factors that might hinder the organization from proceeding with the adoption process. These factors are referred to as discontinuing factors.

If no discontinuing factors are found the process framework enters into its Second stage. The objective of Stage 2 is to determine, from a project –level perspective, the extent to which agile practices can be adopted. The process framework considers each project that wants to adopt Agile practices and assesses the pertinent characteristics of that project that can influence the degree to which it can adopt agile practices. The highest level of agility a project can seek to adopt is referred to as the Project's Potential Agile Level (PPAL). The PPAL is a number from 1 to 5 corresponding to the 5 Agile Levels defined by this framework.

Once the PPAL is identified, the process framework enters its final stage (Stage 3). The objective of Stage 3 is to assess the extent to which an organization possesses the characteristics that will support the adoption of each of the agile practices identified by the PPAL. The outcome of the organizational assessment is the identification of those characteristics needed for the successful adoption of each of the Agile practices and the extent to which those characteristics are present (or absent) in the organization.

An analysis of the "gap" between the project–level and organizational agility assessments indicate the probability of a successful (or unsuccessful) adoption of the agile practices for the project under consideration. If the gap analysis shows a small distance between the PPAL and the organization's readiness then there is a high probability of success for that project to adopt the agile practices of its identified Agile level. If there is a large gap then the project can choose to lower its agile level aspirations, i.e. choose a lower PPAL, and then repeat the gap analysis to ensure that the organization is ready for the new agile level. Another alternative when there is a large gap is to introduce the necessary organizational characteristic(s) to reduce the gap between the PPAL and the organizational readiness. This alternative is conditional on the organization's willingness and ability to go through organizational change. If such changes are implemented the organizational assessment is repeated to identify the new gap distance, and subsequently, the probability of successfully achieving a project's potential Agile level.

These are the basic 3 stages of the process framework. The next sections of this document present the detailed artifacts and assessment documents for each of the stages of this process framework.

# 1. Stage 1: Identifying Discontinuing Factors

## *1.1. General Introduction*

The first stage of the process framework is to identify any factors within the organization that may inhibit the process of adopting Agile practices. This stage is crucial at the beginning of the process due to the fact that it protects the organization from incurring additional cost or effort in trying to adopt Agile practices while "showstoppers" exist that prevent the success of that endeavor. In the process framework these inhibiting factors are referred to as "Discontinuing Factors". Discontinuing factors can vary between different organizations; however, this research has identified four main factors that qualify as discontinuing factors in most situations. If there is a high degree of existence of any of the discontinuing factors then it is recommended that the organization suspends the Agile adoption process until these discontinuing factors are resolved. The four discontinuing factors identified by the process framework are:

1. No value added from adopting agile software development
2. No Executive buy-in to move to agile software development
3. No Finances available to support the transition to agile software development
4. The software being developed is either Mission/Life Critical

The process framework presents a method by which the degree of presence or absence of each of these discontinuing factors in an organization can be measured. The basic concept behind the assessment approach is identifying a group of measurable characteristics for each of the discontinuing factors. Each of these characteristics is measured using a set of indicators.

In the process framework each discontinuing factor has an associated assessment table. The assessment table contains the list of characteristics that need to be measured in order to determine the extent to which the discontinuing factors are present or absent. Each characteristic in the table is coupled with a set of indicators that can be used for the actual assessment process. For example, a portion of the assessment table associated with the first discontinuing factor is illustrated by Table 1 below.

## Will there be any value added from adopting agile software development

| Category of Assessment | Area to be assessed | Characteristic(s) to be assessed | To determine: | Assessment Method | Sample Indicators |
|---|---|---|---|---|---|
| Organization | Project History | Schedule and Budget | Whether or not the organization has a trend of having projects that go over time or over budget | Observation | DC_A1, DC_A2 |
| | Software Process | Problems | Whether the organization is facing any problems or displeasures with the current software process | Interviewing | DC_D1, DC_D2, DC_D3, DC_M1, DC_M2, DC_M3 |

**Table 1. Sample of the assessment table for a discontinuing factor**

The above assessment table can be interpreted by reading it from left to right. For example:

In order to assess whether **there is any value added from adopting agile software development or not,** the assessor needs to examine the **organization itself. (***The only other category that can be examined for discontinuing factors is the project itself).* More specifically in the organization the area of interest to the assessor is **project history.** The assessment focuses on inspecting the particular characteristic of **scheduling and budget** in order to determine **whether or not the organization has a trend of having projects that go over time or budget.** The assessment can be performed by **observation** of the following items (indicators) **DC_A1 and DC_A2.**

It is important to note a couple of things about the “sample indicators” column. These indicators are provided as the means by which to measure what is stated in the “To determine” column. The framework provides a set of suggested indicators, however the assessor performing the actual assessment may add other questions deemed appropriate in order to measure that specific characteristic in their environment.

All the sample indicators in the framework has been placed in the section immediately following the assessment tables. The indicators are grouped based on the person who can best supply the answer to that indicator. This person is identified in the indicator’s name using the first letter after the underscore. For example, if the character after the underscore is a **M** then that sample indicator is found in the table that contains all the questions that should be answered by a manager. In effect:

- A; represents an indicator that needs to be answered by the assessor through observation,
- D; represents an indicator that needs to be answered by a developer, and
- M; represents an indicator that needs to be answered by a manager.

The four discontinuing factors, their assessment tables, and the indicators are provided next.

## 1.2. Assessment Tables for Discontinuing Factors

### **Factor 1:** Will there be any value added from adopting agile software development

| Category of Assessment | Area to be assessed | Characteristic (s) to be assessed | To determine: | Assessment Method | Sample Indicators |
|---|---|---|---|---|---|
| Organization | Project History | Schedule and Budget | Whether or not the organization has a trend of having projects that go over time and budget | Observation | DC_A1, DC_A2 |
| | Software Process | Problems | Whether or not the organization is facing any problems or dissatisfaction with the current software process | Interviewing | DC_D1, DC_D2, DC_D3, DC_M1, DC_M2, DC_M3 |
| Project | Delivery | Time to Market | Whether or not the project has to be developed quickly in order to introduce it to the market as soon as possible | Interviewing | DC_M4 |
| | Requirements | Rate of Change | Whether or not the project's requirements are clear and well defined, thus predicting no change, or whether or not the requirements need to be flexible and/or might change | Interviewing | DC_M5, DC_M6, DC_M7 |

### **Factor 2:** Will there be Executive buy-in to move to agile software development

| Category of Assessment | Area to be assessed | Characteristic(s) to be assessed | To determine: | Assessment Method | Sample Indicators |
|---|---|---|---|---|---|
| People | Managers / Executives | Buy-in | Whether or not executive-level management can see benefits of adopting agile processes and will buy in to the development of agile software | Interviewing | DC_M3, DC_M8, DC_M9 |

### **Factor 3:** Will finances be available to support the transition to agile software development

| Category of Assessment | Area to be assessed | Characteristic(s) to be assessed | To determine: | Assessment Method | Sample Indicators |
|---|---|---|---|---|---|
| Organization | Budget | Availability of Funds | Whether or not the organization has funds to be spent on the adoption process of agile processes and is willing to spend them on the adoption process | Interviewing | DC_M10, DC_M11, DC_M12, DC_M13, DC_M14, DC_M15 |

### **Factor 4:** The software being developed is not Mission or Life Critical

| Category of Assessment | Area to be assessed | Characteristic(s) to be assessed | To determine: | Assessment Method | Sample Indicators |
|---|---|---|---|---|---|
| Project | Criticality | Result of Failure | Whether or not the failure of the software will result in catastrophic losses of lives or failure of missions | Interviewing | DC_M16, DC_M17, DC_M18 |

# 1.3. Indicators

## 1.3.1. Questions to be answered by Managers/Executives

To what extent do you agree with the statements below:

| | Statements | Nominal Values | | | | |
|---|---|---|---|---|---|---|
| | | V | W | X | Y | Z |
| DC_M1 | There are many areas in the development process that always cause problems and/or are inefficient. | Strongly Disagree | Tend to Disagree | Neither Agree nor Disagree | Tend to Agree | Strongly Agree |
| DC_M2 | The current development process is insufficient and/or does not produce good software. | Strongly Disagree | Tend to Disagree | Neither Agree nor Disagree | Tend to Agree | Strongly Agree |
| DC_M3 | There is a need to change the software process in the organization. | Strongly Disagree | Tend to Disagree | Neither Agree nor Disagree | Tend to Agree | Strongly Agree |
| DC_M4 | The customer/client needs to introduce the product to the market quickly. (short time to market). | Strongly Disagree | Tend to Disagree | Neither Agree nor Disagree | Tend to Agree | Strongly Agree |
| DC_M5 | There is a high probability that requirements will change during the development process. | Strongly Disagree | Tend to Disagree | Neither Agree nor Disagree | Tend to Agree | Strongly Agree |
| DC_M6 | Not all the requirements will be known before development starts for the project. | Strongly Disagree | Tend to Disagree | Neither Agree nor Disagree | Tend to Agree | Strongly Agree |
| DC_M7 | The deliverables for this project are unknown. | Strongly Disagree | Tend to Disagree | Neither Agree nor Disagree | Tend to Agree | Strongly Agree |
| DC_M8 | In general, employing agile processes help organizations overcome their software development challenges and/or respond better to customer requests. | Strongly Disagree | Tend to Disagree | Neither Agree nor Disagree | Tend to Agree | Strongly Agree |
| DC_M9 | An Agile Development approach is ideal for the upcoming project. | Strongly Disagree | Tend to Disagree | Neither Agree nor Disagree | Tend to Agree | Strongly Agree |
| DC_M10 | The organization has money allocated for training. | Strongly Disagree | Tend to Disagree | Neither Agree nor Disagree | Tend to Agree | Strongly Agree |
| DC_M11 | The organization has money allocated for process improvement and/or organizational change. | Strongly Disagree | Tend to Disagree | Neither Agree nor Disagree | Tend to Agree | Strongly Agree |
| DC_M12 | The organization is willing to spend on training people about Agile Processes. | Strongly Disagree | Tend to Disagree | Neither Agree nor Disagree | Tend to Agree | Strongly Agree |
| DC_M13 | The organization is willing to spend whatever it takes for project to adopt an Agile Development approach. | Strongly Disagree | Tend to Disagree | Neither Agree nor Disagree | Tend to Agree | Strongly Agree |
| DC_M14 | The organization has the necessary funds to undergo the process of adopting an agile development approach for the upcoming project. | Strongly Disagree | Tend to Disagree | Neither Agree nor Disagree | Tend to Agree | Strongly Agree |
| DC_M15 | If adopting an agile process means buying new software, the organization is able and ready to spend on such software. | Strongly Disagree | Tend to Disagree | Neither Agree nor Disagree | Tend to Agree | Strongly Agree |
| DC_M16 | A failure in this software would not result in the loss of any lives. | Strongly Disagree | Tend to Disagree | Neither Agree nor Disagree | Tend to Agree | Strongly Agree |
| DC_M17 | A failure in this software would not result in the failure of a mission; there is a backup system (there is recourse). | Strongly Disagree | Tend to Disagree | Neither Agree nor Disagree | Tend to Agree | Strongly Agree |
| DC_M18 | The software component being built is not a critical component in a mission or life critical system. | Strongly Disagree | Tend to Disagree | Neither Agree nor Disagree | Tend to Agree | Strongly Agree |

### *1.3.2. Questions to be answered by Developers*

| | Statements | Nominal Values | | | | |
|---|---|---|---|---|---|---|
| | | V | W | X | Y | Z |
| DC_D1 | There are many areas in the development process that always cause problems and/or are inefficient. | Strongly Disagree | Tend to Disagree | Neither Agree nor Disagree | Tend to Agree | Strongly Agree |
| DC_D2 | The current development process is insufficient and/or does not produce good software. | Strongly Disagree | Tend to Disagree | Neither Agree nor Disagree | Tend to Agree | Strongly Agree |
| DC_D3 | There is a need to change the software process in the organization. | Strongly Disagree | Tend to Disagree | Neither Agree nor Disagree | Tend to Agree | Strongly Agree |

### *1.3.3. Questions to be answered by the Assessor through Observation*

| | Statements | Nominal Values | | | | |
|---|---|---|---|---|---|---|
| | | V | W | X | Y | Z |
| DC_A1 | It can be concluded from the previous project plans and the project delivery documents that the organization has been on-time when delivering its projects. | Strongly Disagree | Tend to Disagree | Neither Agree nor Disagree | Tend to Agree | Strongly Agree |
| DC_A2 | It can be concluded from previous project estimates and the project delivery documents that the organization has been within budget for its delivered projects. | Strongly Disagree | Tend to Disagree | Neither Agree nor Disagree | Tend to Agree | Strongly Agree |

# 2. Stage 2: Project-Level Assessment

## *2.1. General Introduction*

Project Level Assessment is responsible for determining the Project's Potential Agile Level (PPAL). The Agile Potential level of a project is that highest level of agility that project can adopt and is denoted by one of the agile levels from the Agile Measurement index (See Appendix A). The highest level of agility a project can adopt behaves as the reference point against which the organizational readiness assessment (Stage 3) will be conducted to discover whether or not the organization is prepared to successful adopt the agile practices and concepts of that agile level.

The project's Agile Potential level is established by assessing the extent to which project level characteristics are present or absent. A project may have numerous characteristics to be assessed, however the ones that are assessed for determining the project agile potential are those that are needed for the successfully adoption a particular group of agile practices and concepts from the 5 Agile levels. This special group of agile practices and concepts are selected *based on the fact that their adoption is dependant on the presence of project characteristics that are outside the control of the project and organization*. Since the project or organization are unable to change the extent to which these project characteristics are present, these project characteristics constrain the level of agility the project can adopt and hence become responsible for determining the project's agile potential.

For example, *frequent face-to-face communication* is a desired agile practice. A project characteristic needed to successfully adopt this practice is *near team proximity*. Assume that the project and organization have no say in changing this project characteristic (*near team proximity*). In other words, change the team's proximity is something outside of their control. If the Project level assessment determines that project characteristic (*near team proximity*) is absent then the highest level of agility for this project will be the same level of agility this agile practice is found in. In the agile levels defined by this research (Appendix A), *frequent face-to-face communication* is an agile practice in level 3. Therefore, in this example the highest level of agility for this project would also be level 3.

In summary, the highest level of agility is determined once the assessment discovers that one of the project characteristics needed to adopt an agile practice or concept are absent, and the project or organization can not do anything to influence or change the absence of that project characteristic.

Table 2 below highlights a subset of agile practices from all the agile practices in the 5 Agile Levels (Appendix A). The practices were chosen in particular because the successful adoption of any of these practices or concepts relies on project characteristic that are outside the control of the organization.

| Agile Level | Agile Principles | | |
|---|---|---|---|
| | **Human Centric** | **Technical Excellence** | **Customer Collaboration** |
| **Level 1 : Cooperative** | | | Customer Commitment to work with Developing Team |
| **Level 2 : Evolutionary** | | | Customer Contract reflective of Evolutionary Development |
| **Level 3 : Effective** | Frequent face-to-face communication between the team | Have around 30% of Level 2 and Level 3 people on team | |
| **Level 4 : Adaptive** | | | Collaborative, Representative, Authorized, Committed, Knowledgeable (CRACK) Customer Immediately Accessible<br>Customer contract revolves around commitment of collaboration, not features |
| **Level 5 : Ambient** | Ideal Agile Physical Setup (The team is in the same room, no cubicles) | No/Minimal number of Level -1 or 1b People on team | Frequent Face-to-face interaction between developers & Users (Collocated) |

**Table 2. Constraining Agile practices and concepts**

The rational behind having this stage as part of the process framework is to minimize the effort and time involved for the third stage of the process framework; assessing how ready the organization is to adopt all the agile practices and concepts. By determining the highest level of agility a project can adopt the need to assess the readiness of the organization for all the agile practices and concepts is eliminated. Instead there is only the need to assess the organizational readiness for the agile practices and concepts of the highest agile level the project can adopt and the ones under that; hence saving assessment time and money.

After briefly describing this stage, the next step is to illustrate how to measure whether a constraining practices or concepts can be adopted in a particular project or not. The Process framework presents a method to conduct that assessment. An assessment method of either interviewing or observation will be assigned to each of the constraining practices or concepts. Based on the selected assessment method, a suggestive set of questions or observations will be provided. All this information is put into an assessment table as depicted by **Error! Reference source not found.** below. The Agile practice or concept being assessed is written under the “To determine” column.

The assessment tables for the constraining agile practices and concepts, and their indicators are provided next.

## 2.2. *Assessment tables for Project Level Agile Constraining Practices*

### Factor to Assess for PPAL to be Level 1 (Assessment Level A):

| Agile Principle to be assessed | To determine: | Assessment Method | Sample Indicators |
|---|---|---|---|
| Customer Collaboration | Whether or not the customer is committed to work with Developing Team | Interviewing | AC_C1, AC_C2, AC_C3, AC_C4 |

### Factor to Assess for PPAL to be Level 2 (Assessment Level B):

| Agile Principle to be assessed | To determine: | Assessment Method | Sample Indicators |
|---|---|---|---|
| Customer Collaboration | Whether or not the customer contract can be reflective of evolutionary development | Interviewing | AC_C5, AC_C6, AC_C7, AC_C8 |

### Factors to Assess for PPAL to be Level 3 (Assessment Level C):

| Agile Principle to be assessed | To determine: | Assessment Method | Sample Indicators |
|---|---|---|---|
| Technical Excellence | Whether or not the development team has around 30% of Level 2 and Level 3 people on team | Observation | AC_A5 |
| Human Centric | Whether or not frequent face-to-face communication between team members is achievable | Observation | AC_A1, AC_A2, AC_A3, AC_A4 |

## Factors to Assess for PPAL to be Level 4 (Assessment Level D):

| Agile Principle to be assessed | To determine: | Assessment Method | Sample Indicators |
|---|---|---|---|
| Customer Collaboration | Whether or not a CRACK customer can be immediately accessible | Interviewing | AC_C9, AC_C10, AC_C11, AC_C12 AC_C13 |
| | Whether or not the customer's contract can revolve around commitment of collaboration, not features | Interviewing | AC_C14, AC_C15, AC_C16, AC_C17 |

## Factors to Assess for PPAL to be Level 5 (Assessment Level E):

| Agile Principle to be assessed | To determine: | Assessment Method | Sample Indicators |
|---|---|---|---|
| Customer Collaboration | Whether or not the frequent face-to-face interaction between developers and customer is achievable | Interviewing | AC_C18, AC_C19 |
| Technical Excellence | Whether or not no or a minimal number of Level -1 or 1b people exists on the development team | Observation | AC_A6 |
| Human Centric | Whether or not it is feasible to have an ideal agile physical setup | Observation | AC_A7, AC_A8, AC_A9 |

# 2.3. Indicators

## 2.3.1. Questions to be answered by the Client/Customer

To what extent do you agree with the statements below:

| | Statements | Nominal Values | | | | |
|---|---|---|---|---|---|---|
| | | V | W | X | Y | Z |
| AC_C1 | The customer is willing to dedicate time to take an active role in this project. | Strongly Disagree | Tend to Disagree | Neither Agree nor Disagree | Tend to Agree | Strongly Agree |
| AC_C2 | In the past, the customer has dedicated time to collaborate with the development team. | Strongly Disagree | Tend to Disagree | Neither Agree nor Disagree | Tend to Agree | Strongly Agree |
| AC_C3 | The customer believes that the contractor should make most of the effort and that the customer should have to do little other than check on the project's status and do a final acceptance. | Strongly Disagree | Tend to Disagree | Neither Agree nor Disagree | Tend to Agree | Strongly Agree |
| AC_C4 | The customer is committed to working with the development team. | Strongly Disagree | Tend to Disagree | Neither Agree nor Disagree | Tend to Agree | Strongly Agree |
| AC_C5 | The customer agrees to have the system developed in an iterative/incremental fashion as opposed to the approach of a big delivery at the end of the contracted time. | Strongly Disagree | Tend to Disagree | Neither Agree nor Disagree | Tend to Agree | Strongly Agree |
| AC_C6 | The customer is willing to sign a contract to start development of a product whose requirements cannot be known ahead of time with certainty. | Strongly Disagree | Tend to Disagree | Neither Agree nor Disagree | Tend to Agree | Strongly Agree |
| AC_C7 | The customer is willing to change its typical contract structure to reflect an evolutionary development approach. Evolutionary development implies that the requirements, plan, estimates, and solution evolve or are refined over the course of the iterations, instead of being fully defined and "frozen" in a major upfront specification effort before the development begins. | Strongly Disagree | Tend to Disagree | Neither Agree nor Disagree | Tend to Agree | Strongly Agree |
| AC_C8 | The customer is willing to accept an overall project plan and a detailed plan of the next iteration only. The customer does not have a problem with not receiving a GANTT or PERT chart of the whole project upfront. | Strongly Disagree | Tend to Disagree | Neither Agree nor Disagree | Tend to Agree | Strongly Agree |
| AC_C9 | The customer representative(s) interacting with the contracted organization is (are) authorized to make decisions on the spot regarding the product specifications | Strongly Disagree | Tend to Disagree | Neither Agree nor Disagree | Tend to Agree | Strongly Agree |
| AC_C10 | The customer representative(s) interacting with the contracted organization is (are) knowledgeable about the product domain (i.e. he/she is a domain expert or subject matter expert). | Strongly Disagree | Tend to Disagree | Neither Agree nor Disagree | Tend to Agree | Strongly Agree |
| AC_C11 | The customer representative(s) interacting with the contracted organization is (are) representative of the product's actual users. | Strongly Disagree | Tend to Disagree | Neither Agree nor Disagree | Tend to Agree | Strongly Agree |
| AC_C12 | The customer representative is available for the development team to contact if it needs his/her input on something. | Strongly Disagree | Tend to Disagree | Neither Agree nor Disagree | Tend to Agree | Strongly Agree |
| AC_C13 | The customer representative is immediately accessible to the development team if needed. | Strongly Disagree | Tend to Disagree | Neither Agree nor Disagree | Tend to Agree | Strongly Agree |
| AC_C14 | The customer is willing to sign a contract that does not have a detailed enumeration of features and functions but broad goals and the success criteria. This allows the customer more flexibility to change and add requirements through out the development process. | Strongly Disagree | Tend to Disagree | Neither Agree nor Disagree | Tend to Agree | Strongly Agree |
| AC_C15 | The customer is willing to accept a contract in which the time and budget, but not the features to be delivered, are fixed. | Strongly Disagree | Tend to Disagree | Neither Agree nor Disagree | Tend to Agree | Strongly Agree |
| AC_C16 | The customer is willing to accept a contract that commits both sides to a degree of interaction and collaboration instead of a set of detailed requirements. | Strongly Disagree | Tend to Disagree | Neither Agree nor Disagree | Tend to Agree | Strongly Agree |
| AC_C17 | The customer is willing to change its typical contract structure to reflect a new agile development approach. An agile development approach will give the customer the flexibility to change its requirements throughout the development process, and will deliver software earlier and in increments. | Strongly Disagree | Tend to Disagree | Neither Agree nor Disagree | Tend to Agree | Strongly Agree |
| AC_C18 | The customer will be available for frequent face-to-face interaction with the development team. | Strongly Disagree | Tend to Disagree | Neither Agree nor Disagree | Tend to Agree | Strongly Agree |
| AC_C19 | The customer is willing to be collocated with the development team. | Strongly Disagree | Tend to Disagree | Neither Agree nor Disagree | Tend to Agree | Strongly Agree |

# *2.3.2. Questions to be answered by the Assessor through Observation*

To what extent do you agree with the statements below:

| | Statements | Nominal Values | | | | |
|---|---|---|---|---|---|---|
| | | V | W | X | Y | Z |
| AC_A1 | The development team is located where members can have frequent face-to-face communication. | Strongly Disagree | Tend to Disagree | Neither Agree nor Disagree | Tend to Agree | Strongly Agree |
| AC_A2 | The geographic distribution of the development team can be best described as… | Within Flying Distance | Within driving distance | Within the same city/area | Within the same building | In the same room |
| AC_A3 | Logistically, the development team can meet face-to-face. | Yearly or never | Monthly | Weekly | Daily | Hourly |
| AC_A4 | It is likely for the development team to have frequent face-to-face communication. | Strongly Disagree | Tend to Disagree | Neither Agree nor Disagree | Tend to Agree | Strongly Agree |
| AC_A5 | What percentage of the full-time staff is of Cockburn Level 2 or Level 3 experts | 0-5% | 5-10% | 10-15% | 15-30% | 30% or higher |
| AC_A6 | Indicate the percentage of full-time staff who are Cockburn Level 2 or Level 3 experts. | 30% or higher | 15-30% | 10-15% | 5-10% | 0-5% |
| AC_A7 | It is highly probable that the organization can have all the development personnel in a common room rather than separate offices or cubicles. | Strongly Disagree | Tend to Disagree | Neither Agree nor Disagree | Tend to Agree | Strongly Agree |
| AC_A8 | It is highly probable that the organization can set up the development rooms to better support agile development (furniture away from the walls). | Strongly Disagree | Tend to Disagree | Neither Agree nor Disagree | Tend to Agree | Strongly Agree |
| AC_A9 | It is highly probable that the organization can setup an environment where as much project information as possible is displayed on the walls (via whiteboards, cling sheets, or projectors). | Strongly Disagree | Tend to Disagree | Neither Agree nor Disagree | Tend to Agree | Strongly Agree |

# 3. Stage 3: Organizational Readiness

After the PPAL is determined from stage 2, the process framework advances to the third and final stage; assessing the organization's readiness to adopt the agile practices and concepts that the project aims to adopt. During this assessment of readiness the organization is assessed from various perspectives such as culture, environments, process, management, developers and various other perspectives depending on the agile practice to be adopted. For example, assume a certain Project's Potential Agile Level is determined to be 3 (PPAL =3). In that case we need to assess the organization's readiness to adopt each of the agile practices and concepts defined in level 1, 2 and 3. The approach used to assess the organizational readiness for an agile practice is similar to the method used for the assessment of the discontinuing factors. Let's take a look at a quick example. The example below illustrates how to assess the organization's readiness to adopt an agile practice known as "Coding Standards".

## Coding Standards

| Category of Assessment | Area to be assessed | Characteristic(s) to be assessed | To determine: | Assessment Method | Sample Indicators |
|---|---|---|---|---|---|
| People | Developers | Buy-In | Whether the developers see the benefit and are willing to apply coding standards | Interviewing | OR1_D21, OR1_D22 |
| Process | Coding Standards | Existence | Whether there exists any kind of coding standards that are used | Observation | OR1_A2 |

**Table 3. Assessment table used to measure organizational readiness for agile practices**

Table 3, above shows that in order to assess the organization's readiness the assessor first looks at the "people" in the organization with a particular focus on the developers. The characteristic that needs to be assessed with regards to the developers is their buy-in. The assessment of the developers' buy-in will help to determine whether or not the developers can recognize the benefits of coding standards and are willing to adopt them. The 5th column in the assessment table suggests a method of the assessment process; in this case it is interviewing. The process framework also provides a suggestive list of sample indicators (or questions) that can be used for the assessment process. The assessor is free to use other assessment methods or indicators as long as the characteristics that need to be assessing are validly measured. In the example above, for the assessment to be complete the "process" also needs to be assessed. Within the process the focus is on the existence of current coding standards. The method advised for this assessment is observation, and the sample indicator column will highlight what exactly needs to be observed.

Following this brief introduction of stage 3 are 5 subsections, each one related to the assessing the organizational readiness for of the agile practices and concepts found in the one of the Agile Levels.

## *3.1. Assessment Tables for Practices and Concepts in Agile Level 1*

## Collaborative Planning (*Customers, Developers and Management plan together)*

| Category of Assessment | Area to be assessed | Characteristic(s) to be assessed | To determine: | Assessment Method | Sample Indicators |
|---|---|---|---|---|---|
| People | Management | Management Style | Whether or not a collaborative or a command-control relation exists between managers and subordinates. [12, 10]. The management style is an indication of whether or not management trusts the developers and vice-versa. | Interviewing | OR1_M1, OR1_M2, OR1_M3, OR1_M4, OR1_M5, OR1_M14, OR1_M17, OR1_D1 OR1_D2, OR1_D3, OR1_D4, |
| | | Buy-In | Whether or not management is supportive of or resistive to having a collaborative environment | Interviewing | OR1_M9, OR1_M10, |
| | | Transparency | Whether or not management can be open with customers and developers – No politics and secrets [16] | Interviewing | OR1_M6, OR1_M7, OR1_M8, OR1_M13 |
| | Developers | Power Distance | Whether or not people are intimidated/afraid to give honest feedback and participation in the presence of their managers | Interviewing | OR1_M11, OR1_D6, OR1_D7, OR1_D8, OR1_D9 |
| | | Buy-In | Whether or not the developers are willing to plan in a collaborative environment | Interviewing | OR1_D5 |
| Project Management | Planning | Existence | Whether or not the organization does basic planning for its projects | Observation | OR1_A1 |
| | | | | Interviewing | OR1_M16, OR1_M18 |

## Task Volunteering not Task Assignment

| Category of Assessment | Area to be assessed | Characteristic(s) to be assessed | To determine: | Assessment Method | Sample Indicators |
|---|---|---|---|---|---|
| People | Management | Buy-In | Whether or not management will be willing to buy into and can see benefits from employees volunteering for tasks instead of being assigned | Interviewing | OR1_M12, OR1_M15 |
| | Developers | Buy-In | Whether or not developers are willing to see the benefits from volunteering for tasks | Interviewing | OR1_D10 |

## Collaborative Teams

| Category of Assessment | Area to be assessed | Characteristic(s) to be assessed | To determine: | Assessment Method | Sample Indicators |
|---|---|---|---|---|---|
| People | Developers | Interaction | Whether or not any levels of interaction exist between people thus laying a foundation for more team work | Interviewing | OR1_M1, OR1_D15 |
| | | Collectivism | Whether or not people believe in group work and helping others or are just concerned about themselves | Interviewing | OR1_D16 |
| | | Buy-In | Whether or not people are willing to work in teams | Interviewing | OR1_D12, OR1_D11 |
| | | | Whether or not people recognize that their input is valuable in group work | Interviewing | OR1_D23, OR1_D13 |

## Empowered and Motivated Teams [5]

| Category of Assessment | Area to be assessed | Characteristic(s) to be assessed | To determine: | Assessment Method | Sample Indicators |
|---|---|---|---|---|---|
| People | Developers | Decision Making | Whether or not management empowers teams with decision making authority | Interviewing | OR1_M3, OR1_D4, OR1_D14, OR1_D17, OR1_M14 |
| | | Motivation | Whether or not people are treated in a way that motivates them | Interviewing | OR1_D14, OR1_D13, OR1_D23, OR1_D24, OR1_D25, OR1_D15 |
| | Managers | Trust | Whether or not managers trust and believe in the technical team in order to truly empower them | Interviewing | OR1_M13, OR1_M14, OR1_M6, OR1_M12, OR1_D2 |

## Coding Standards [11, 19, 15]

| Category of Assessment | Area to be assessed | Characteristic(s) to be assessed | To determine: | Assessment Method | Sample Indicators |
|---|---|---|---|---|---|
| People | Developers | Buy-In | Whether or not the developers see the benefit and are willing to apply coding standards | Interviewing | OR1_D21, OR1_D22 |
| Process | Coding Standards | Existence | Whether or not any kind of coding standards exists that are used | Observation | OR1_A2 |

## Knowledge Sharing [13]

| Category of Assessment | Area to be assessed | Characteristic(s) to be assessed | To determine: | Assessment Method | Sample Indicators |
|---|---|---|---|---|---|
| People | Developers | Buy-In | Whether or not developers believe in and can see the benefits of having project information communicated to the whole team | Interviewing | OR1_D18, OR1_D19, OR1_M19 |
| | Managers | Buy-In | Whether or not managers believe in and can see the benefits of having project information communicated to the whole team | Interviewing | OR1_M6, OR1_M7, OR1_M20, OR1_M21, OR1_M22 |
| Tools | Knowledge Sharing | Availability | Whether or not knowledge sharing tools are available and accessible (Wikis, Blogs …etc.) | Observation | OR1_A3 |

## Reflect and Tune Process

| Category of Assessment | Area to be assessed | Characteristic(s) to be assessed | To determine: | Assessment Method | Sample Indicators |
|---|---|---|---|---|---|
| People | Developers | Buy-in | Whether or not developers are willing to commit to reflecting about and tuning the process after each iteration or release | Interviewing | OR1_D26 |
| | Managers | Buy-in | Whether or not management is willing to commit to reflecting about and tuning the process after each iteration or release | Interviewing | OR1_M23 |
| Process | Process improvement | Capability | Whether or not the organization can handle process change in the middle of the project | Interviewing | OR1_D27, OR1_D28, OR1_D29, OR1_M24, OR1_M25, OR1_M26 |

# *3.1.1. Indicators for Agile Level 1*

## *3.1.1.1. Questions to be answered by Developers*

To what extent do you agree with the statements below:

| | Statements | Nominal Values | | | | |
|---|---|---|---|---|---|---|
| | | V | W | X | Y | Z |
| OR1_D1 | Your manager is collaborative. | Strongly Disagree | Tend to Disagree | Neither Agree nor Disagree | Tend to Agree | Strongly Agree |
| OR1_D2 | Your manager does not micro-manage you or your work. | Strongly Disagree | Tend to Disagree | Neither Agree nor Disagree | Tend to Agree | Strongly Agree |
| OR1_D3 | Your manager encourages you to be creative and does not dictate to you what to do exactly. | Strongly Disagree | Tend to Disagree | Neither Agree nor Disagree | Tend to Agree | Strongly Agree |
| OR1_D4 | Your manager gives you the authority to make decisions without referring back to him/her. | Strongly Disagree | Tend to Disagree | Neither Agree nor Disagree | Tend to Agree | Strongly Agree |
| OR1_D5 | You would like to participate in the planning process of the project you will work on. | Strongly Disagree | Tend to Disagree | Neither Agree nor Disagree | Tend to Agree | Strongly Agree |
| OR1_D6 | If your manager said or did something wrong, it is acceptable for you to correct and/or constructively criticize him/her face to face. | Strongly Disagree | Tend to Disagree | Neither Agree nor Disagree | Tend to Agree | Strongly Agree |
| OR1_D7 | It is acceptable for you to express disagreement with your manager(s) without fearing their retribution. | Strongly Disagree | Tend to Disagree | Neither Agree nor Disagree | Tend to Agree | Strongly Agree |
| OR1_D8 | In a group meeting, the customer suggested something about the product. You disagree and have a better idea; it is acceptable for you to express disagreement with your customer and suggest something better. | Strongly Disagree | Tend to Disagree | Neither Agree nor Disagree | Tend to Agree | Strongly Agree |
| OR1_D9 | Other peoples' titles and positions intimidate people in the organization. | Strongly Disagree | Tend to Disagree | Neither Agree nor Disagree | Tend to Agree | Strongly Agree |
| OR1_D10 | You would do a better job choosing your own task on a project instead of being assigned one by your manager. | Strongly Disagree | Tend to Disagree | Neither Agree nor Disagree | Tend to Agree | Strongly Agree |
| OR1_D11 | You prefer working in a group. | Strongly Disagree | Tend to Disagree | Neither Agree nor Disagree | Tend to Agree | Strongly Agree |
| OR1_D12 | Indicate how often you work in groups. | Never | Seldom | Sometimes | Usually | Always |
| OR1_D13 | When in a group, you feel that your participation is important. | Strongly Disagree | Tend to Disagree | Neither Agree nor Disagree | Tend to Agree | Strongly Agree |
| OR1_D14 | Your manager seeks your input on technical issues. | Strongly Disagree | Tend to Disagree | Neither Agree nor Disagree | Tend to Agree | Strongly Agree |
| OR1_D15 | Your team members seek your input on technical issues. | Strongly Disagree | Tend to Disagree | Neither Agree nor Disagree | Tend to Agree | Strongly Agree |
| OR1_D16 | When you run into technical problems, you usually ask your team members about the solution. | Strongly Disagree | Tend to Disagree | Neither Agree nor Disagree | Tend to Agree | Strongly Agree |
| OR1_D17 | You usually participate in the planning process of the project you are working on. | Strongly Disagree | Tend to Disagree | Neither Agree nor Disagree | Tend to Agree | Strongly Agree |
| OR1_D18 | Project information should be communicated to the whole team. | Strongly Disagree | Tend to Disagree | Neither Agree nor Disagree | Tend to Agree | Strongly Agree |
| OR1_D19 | There should be a mechanism for persistent knowledge sharing between team members. | Strongly Disagree | Tend to Disagree | Neither Agree nor Disagree | Tend to Agree | Strongly Agree |
| OR1_D20 | People should use a wiki or a blog for knowledge sharing. | Strongly Disagree | Tend to Disagree | Neither Agree nor Disagree | Tend to Agree | Strongly Agree |
| OR1_D21 | There should exist a coding standard for development. | Strongly Disagree | Tend to Disagree | Neither Agree nor Disagree | Tend to Agree | Strongly Agree |
| OR1_D22 | If the organization has a coding standard, then developers should use it when coding, even in crunch time. | Strongly Disagree | Tend to Disagree | Neither Agree nor Disagree | Tend to Agree | Strongly Agree |
| OR1_D23 | The organization values you and your expertise. | Strongly Disagree | Tend to Disagree | Neither Agree nor Disagree | Tend to Agree | Strongly Agree |
| OR1_D24 | Your manager has high expectations of you. | Strongly Disagree | Tend to Disagree | Neither Agree nor Disagree | Tend to Agree | Strongly Agree |
| OR1_D25 | You are motivated by your job. | Strongly Disagree | Tend to Disagree | Neither Agree nor Disagree | Tend to Agree | Strongly Agree |
| OR1_D26 | You are willing to dedicate time after each iteration/release to review how the process could be improved. | Strongly Disagree | Tend to Disagree | Neither Agree nor Disagree | Tend to Agree | Strongly Agree |
| OR1_D27 | You are willing to undergo a process change even if it requires some reworking of already completed work products. | Strongly Disagree | Tend to Disagree | Neither Agree nor Disagree | Tend to Agree | Strongly Agree |
| OR1_D28 | If there is a need for process change, that change should not be considered a burden on the team even if significant process changes have been made previously during the project. | Strongly Disagree | Tend to Disagree | Neither Agree nor Disagree | Tend to Agree | Strongly Agree |
| OR1_D29 | Process change in the middle of the project should not be considered a disruption since the process change is worth the benefit it will bring. | Strongly Disagree | Tend to Disagree | Neither Agree nor Disagree | Tend to Agree | Strongly Agree |

## *3.1.1.2. Questions to be answered by Managers/Executives*

To what extent do you agree with the statements below:

| | Statements | Nominal Values | | | | |
|---|---|---|---|---|---|---|
| | | V | W | X | Y | Z |
| OR1_M1 | You actively encourage interaction among your subordinates. | Strongly Disagree | Tend to Disagree | Neither Agree nor Disagree | Tend to Agree | Strongly Agree |
| OR1_M2 | You prefer team work over individual work. | Strongly Disagree | Tend to Disagree | Neither Agree nor Disagree | Tend to Agree | Strongly Agree |
| OR1_M3 | You usually seek your subordinates' opinions before making a decision. | Strongly Disagree | Tend to Disagree | Neither Agree nor Disagree | Tend to Agree | Strongly Agree |
| OR1_M4 | You frequently brainstorm with your subordinates. | Strongly Disagree | Tend to Disagree | Neither Agree nor Disagree | Tend to Agree | Strongly Agree |
| OR1_M5 | You frequently encourage your subordinates to find creative solutions to problems. | Strongly Disagree | Tend to Disagree | Neither Agree nor Disagree | Tend to Agree | Strongly Agree |
| OR1_M6 | It is important for you to share project management information with your subordinates. | Strongly Disagree | Tend to Disagree | Neither Agree nor Disagree | Tend to Agree | Strongly Agree |
| OR1_M7 | If you are needed and unreachable, at any point in time your subordinates have enough information to update the customer about the exact status of the project. | Strongly Disagree | Tend to Disagree | Neither Agree nor Disagree | Tend to Agree | Strongly Agree |
| OR1_M8 | If a problem occurs that may affect the schedule or requirements of a project, you would update your client right away. | Strongly Disagree | Tend to Disagree | Neither Agree nor Disagree | Tend to Agree | Strongly Agree |
| OR1_M9 | You believe that developers should aid in the planning of a project. | Strongly Disagree | Tend to Disagree | Neither Agree nor Disagree | Tend to Agree | Strongly Agree |
| OR1_M10 | You believe that customers should be part of the planning of a project. | Strongly Disagree | Tend to Disagree | Neither Agree nor Disagree | Tend to Agree | Strongly Agree |
| OR1_M11 | Other peoples' titles and positions intimidate people in the organization. | Strongly Disagree | Tend to Disagree | Neither Agree nor Disagree | Tend to Agree | Strongly Agree |
| OR1_M12 | You would allow your subordinates to choose their own tasks for a project | Strongly Disagree | Tend to Disagree | Neither Agree nor Disagree | Tend to Agree | Strongly Agree |
| OR1_M13 | Its acceptable for your subordinates to have unregulated access to the customer. | Strongly Disagree | Tend to Disagree | Neither Agree nor Disagree | Tend to Agree | Strongly Agree |
| OR1_M14 | You frequently seek the input of your subordinates on technical issues. | Strongly Disagree | Tend to Disagree | Neither Agree nor Disagree | Tend to Agree | Strongly Agree |
| OR1_M15 | You believe that subordinates would perform better and be more effective if they were to choose their own tasks. | Strongly Disagree | Tend to Disagree | Neither Agree nor Disagree | Tend to Agree | Strongly Agree |
| OR1_M16 | You always create plans for a software development project. | Strongly Disagree | Tend to Disagree | Neither Agree nor Disagree | Tend to Agree | Strongly Agree |
| OR1_M17 | You believe that it is important to involve other people while preparing the project plan. | Strongly Disagree | Tend to Disagree | Neither Agree nor Disagree | Tend to Agree | Strongly Agree |
| OR1_M18 | The project plans are always documented. | Strongly Disagree | Tend to Disagree | Neither Agree nor Disagree | Tend to Agree | Strongly Agree |
| OR1_M19 | When you prepare a project plan, it should not include the details of the project from start to end; it should be focused on the next iteration while giving an overview of the overall work | Strongly Disagree | Tend to Disagree | Neither Agree nor Disagree | Tend to Agree | Strongly Agree |
| OR1_M20 | Project information should be communicated to the whole team. | Strongly Disagree | Tend to Disagree | Neither Agree nor Disagree | Tend to Agree | Strongly Agree |
| OR1_M21 | There should be a mechanism for persistent knowledge sharing between team members. | Strongly Disagree | Tend to Disagree | Neither Agree nor Disagree | Tend to Agree | Strongly Agree |
| OR1_M22 | If there was a wiki or a blog set up for knowledge sharing, you believe people would use it. | Strongly Disagree | Tend to Disagree | Neither Agree nor Disagree | Tend to Agree | Strongly Agree |
| OR1_M23 | You are willing to dedicate time after each iteration/release to review how the process could be improved. | Strongly Disagree | Tend to Disagree | Neither Agree nor Disagree | Tend to Agree | Strongly Agree |
| OR1_M24 | You are willing to undergo a process change even if it requires some reworking of already completed work products. | Strongly Disagree | Tend to Disagree | Neither Agree nor Disagree | Tend to Agree | Strongly Agree |
| OR1_M25 | If there is a need for process change, that change should not be considered a burden on the team even if significant process changes have been made previously during the project. | Strongly Disagree | Tend to Disagree | Neither Agree nor Disagree | Tend to Agree | Strongly Agree |
| OR1_M26 | Process change in the middle of the project should not be considered a disruption since the process change is worth the benefit it will bring. | Strongly Disagree | Tend to Disagree | Neither Agree nor Disagree | Tend to Agree | Strongly Agree |

## *3.1.1.3. Questions to be answered by the Assessor through observations*

To what extent do you agree with the statements below:

| | **Statements** | Nominal Values | | | | |
|---|---|---|---|---|---|---|
| | | V | W | X | Y | Z |
| OR1_A1 | Old project documents show that previous projects have a project plan. | Strongly Disagree | Tend to Disagree | Neither Agree nor Disagree | Tend to Agree | Strongly Agree |
| OR1_A2 | A review of documents or other information shows you that the organization has a coding standard. | Strongly Disagree | Tend to Disagree | Neither Agree nor Disagree | Tend to Agree | Strongly Agree |
| OR1_A3 | A review of the tools available for use by the developers shows you that the organization has knowledge sharing tools (Wikis, Blogs ...etc.) available and accessible. | Strongly Disagree | Tend to Disagree | Neither Agree nor Disagree | Tend to Agree | Strongly Agree |

## *3.2. Assessment Tables for Practices and Concepts in Agile Level 2*

## Evolutionary Requirements [14]

| Category of Assessment | Area to be assessed | Characteristic(s) to be assessed | To determine: | Assessment Method | Sample Indicators |
|---|---|---|---|---|---|
| Process | Requirements Engineering | Existence | Whether or not the organization has an institutionalized procedure to gather requirements from its clients | Observation | OR2_A1 |
| | | | | Interviewing | OR2_M, OR2_M2 |
| | | Experience | Whether or not the organization has developed projects using the evolutionary requirements | Interviewing | OR2_D1, OR2_M3 |
| People | Management | Uncertainty Avoidance | Whether or not management accepts and is comfortable with the uncertainty involved with deciding on requirements and features as late as possible | Interviewing | OR2_M4, OR2_M5, OR2_M6 |
| | | Competence | Whether or not the managers can recognize high-level (architecturally influential) requirements and differentiate them from detail requirements | Interviewing | OR2_M7, OR2_M8 |
| | | Buy-In | Whether or not management is willing to accept changes from the customer and that all changes are reversible | Interviewing | OR2_M6, OR2_M9, OR2_M0 |
| | | | Whether or not management is willing to try evolutionary requirements over big upfront requirements gathering | Interviewing | OR2_M1, OR2_M2 |
| | Developers | Uncertainty Avoidance | Whether or not the developers accept and are comfortable with the uncertainty involved with deciding on requirements and features as late as possible | Interviewing | OR2_D2, OR2_D3 |
| | | Buy-In | Whether or not the developers are willing to accept changes from the customer and that all changes are reversible | Interviewing | OR2_D4, OR2_D7, OR2_D8 |
| | | Competence | Whether or not the developers can recognize high-level (architecturally influential) requirements and differentiate them from detail requirements | Interviewing | OR2_D5,OR2_D6 |

## Software Configuration Management [12]

| Category of Assessment | Area to be assessed | Characteristic(s) to be assessed | To determine: | Assessment Method | Sample Indicators |
|---|---|---|---|---|---|
| Environment | Software Tools | Existence | Whether or not the organization has tools for software configuration management | Observation | OR2_A3 |

## Continuous Delivery (Incremental-Iterative development) [14, 13, 11, 4]

| Category of Assessment | Area to be assessed | Characteristic(s) to be assessed | To determine: | Assessment Method | Sample Indicators |
|---|---|---|---|---|---|
| Process | Process Definition | Existence | Whether or not the organization has any process in place for development and is not relying on haphazard and ad-hoc approaches to software development | Observation | OR2_A2 |
| | | | | Interviewing | OR2_D9, OR2_D10, OR2_M13, OR2_M14 |
| | Lifecycle | Experience | Whether or not the organization has previously used an incremental – iterative approach for developing systems | Interviewing | OR2_M15, OR2_M16 OR2_D11, OR2_D12 |
| People | Management | Buy-In | Whether or not management will be willing to use an iterative-incremental development approach | Interviewing | OR2_M17, OR2_M18 |
| | | Stress | Whether or not managers can handle the additional stress of overlooking the delivery of workable iterations every 1-4 weeks | Interviewing | OR2_M19 |
| | | Competence | Whether or not the managers understand the principles of incremental-iterative development | Interviewing | OR2_M20, OR2_M21 |
| | Developers | Stress | Whether or not the developers can handle the stress of delivering a workable iteration every 1-4 weeks | Interviewing | D_15 |
| | | Buy-In | Whether or not developers will be willing to use an iterative-incremental development approach | Interviewing | OR2_D13, OR2_D14, OR2_D18 |
| | | Competence | Whether or not the developers understand the principles of incremental-iterative development | Interviewing | OR2_D16, OR2_D17 |

## Planning at different levels

| Category of Assessment | Area to be assessed | Characteristic(s) to be assessed | To determine: | Assessment Method | Sample Indicators |
|---|---|---|---|---|---|
| People | Managers | Competence | Whether or not management understands the principles and significance of multi-level planning | Interviewing | OR2_M22, OR2_M23, OR2_M24, OR2_M25, OR2_M26 |
| | | Buy-in | Whether or not management is willing to commit to the process of continuously planning versus developing a one-time plan | Interviewing | OR2_M27, OR2_M28 |
| Process | Planning | Experience | Whether or not the organization is experienced with multi-level or not | Interviewing | OR2_M29 |

## Tracking Iteration Progress through Working Software [13]

| Category of Assessment | Area to be assessed | Characteristic(s) to be assessed | To determine: | Assessment Method | Sample Indicators |
|---|---|---|---|---|---|
| Process | Process Management | Monitoring & Reporting | Whether or not a mechanism exists to monitor the iteration progress is monitored | Interviewing | OR2_M30, OR2_D19, OR2_D21, OR2_M31 |
| People | Managers | Buy-in | Whether or not the managers can see that working software is a valid progress indicator | Interviewing | OR2_M32 |
| | Developers | Buy-in | Whether or not the developers can see that working software is a valid progress indicator | Interviewing | OR2_D20 |

## No Big Design up Front (BDUF)

| Category of Assessment | Area to be assessed | Characteristic(s) to be assessed | To determine: | Assessment Method | Sample Indicators |
|---|---|---|---|---|---|
| Process | Design | Experience | Whether or not design is a continuous process, or done once at the beginning of the development process | Interviewing | OR2_M36, OR2_M37 |
| People | Developers | Buy-in | Whether or not the developers agree to the fact that no big design up front is a valid and efficient approach for agile development | Interviewing | OR2_D22, OR2_D23, OR2_D24 |
| | Managers | Buy-in | Whether managers agree to the fact that no big design up front is a valid and efficient approach for agile development | Interviewing | OR2_M33, OR2_M34, OR2_M35 |

# 3.2.1. Indicators for Agile Level 2

# 3.2.1.1. Questions to be answered by Developers

To what extent do you agree with the statements below:

| | Statements | Nominal Values | | | | |
|---|---|---|---|---|---|---|
| | | V | W | X | Y | Z |
| OR2_D1 | Indicate how often are you involved in a project in which all the requirements are not known upfront and an evolutionary requirements approach is used. | Never | Seldom | Sometimes | Usually | Always |
| OR2_D2 | You can start a development of a project without knowing the exact requirements of the whole project. | Strongly Disagree | Tend to Disagree | Neither Agree nor Disagree | Tend to Agree | Strongly Agree |
| OR2_D3 | If circumstances dictate that all the details are not available before you start a project, you do not mind the uncertainty and floating targets. | Strongly Disagree | Tend to Disagree | Neither Agree nor Disagree | Tend to Agree | Strongly Agree |
| OR2_D4 | You do not mind starting a project knowing that its requirements will evolve or change in the future. | Strongly Disagree | Tend to Disagree | Neither Agree nor Disagree | Tend to Agree | Strongly Agree |
| OR2_D5 | You can tell the difference between requirements that will the influence the architecture and design of a project and requirements that will not influence it. | Strongly Disagree | Tend to Disagree | Neither Agree nor Disagree | Tend to Agree | Strongly Agree |
| OR2_D6 | In a project, you can recognize high level requirements that most probably will not change versus low level requirements that might change. | Strongly Disagree | Tend to Disagree | Neither Agree nor Disagree | Tend to Agree | Strongly Agree |
| OR2_D7 | Throughout the project the client has full right to change the requirements in order to meet his/her business needs. | Strongly Disagree | Tend to Disagree | Neither Agree nor Disagree | Tend to Agree | Strongly Agree |
| OR2_D8 | In order to deliver valuable software to clients, change should be welcomed but not constrained. | Strongly Disagree | Tend to Disagree | Neither Agree nor Disagree | Tend to Agree | Strongly Agree |
| OR2_D9 | Software development in this organization is not ad hoc or haphazard; there is a clear and known process in place. | Strongly Disagree | Tend to Disagree | Neither Agree nor Disagree | Tend to Agree | Strongly Agree |
| OR2_D10 | Every project involves a clear set of activities. Each of these activities has clear standardized deliverables. | Strongly Disagree | Tend to Disagree | Neither Agree nor Disagree | Tend to Agree | Strongly Agree |
| OR2_D11 | Indicate how often you have worked on a project that was developed in an incremental –iterative approach. | Never | Seldom | Sometimes | Usually | Always |
| OR2_D12 | It is a common practice to divide the system up into mini-projects. The system is seldom developed as one large project. | Strongly Disagree | Tend to Disagree | Neither Agree nor Disagree | Tend to Agree | Strongly Agree |
| OR2_D13 | The incremental-iterative approach has more benefits than the waterfall approach. | Strongly Disagree | Tend to Disagree | Neither Agree nor Disagree | Tend to Agree | Strongly Agree |
| OR2_D14 | You are willing to use the incremental-iterative approach to develop software. | Strongly Disagree | Tend to Disagree | Neither Agree nor Disagree | Tend to Agree | Strongly Agree |
| OR2_D15 | Delivering a working increment every 1-4 weeks will not cause you any additional stress. | Strongly Disagree | Tend to Disagree | Neither Agree nor Disagree | Tend to Agree | Strongly Agree |
| OR2_D16 | No big upfront analysis should be conducted when using the incremental-iterative approach. | Strongly Disagree | Tend to Disagree | Neither Agree nor Disagree | Tend to Agree | Strongly Agree |
| OR2_D17 | You fully understand the principles of the incremental-iterative development approach. | Strongly Disagree | Tend to Disagree | Neither Agree nor Disagree | Tend to Agree | Strongly Agree |
| OR2_D18 | You are willing to do more integration (integrate after each iteration) in order to accommodate the incremental-iterative development approach. | Strongly Disagree | Tend to Disagree | Neither Agree nor Disagree | Tend to Agree | Strongly Agree |
| OR2_D19 | The organization has a usable and efficient method for reporting project status. | Strongly Disagree | Tend to Disagree | Neither Agree nor Disagree | Tend to Agree | Strongly Agree |
| OR2_D20 | Working software should be the primary measure of progress in a project. | Strongly Disagree | Tend to Disagree | Neither Agree nor Disagree | Tend to Agree | Strongly Agree |
| OR2_D21 | During development you deliver a software iteration/release at least once within the organizational status-reporting window (usually one month). | Strongly Disagree | Tend to Disagree | Neither Agree nor Disagree | Tend to Agree | Strongly Agree |
| OR2_D22 | Development of the first iteration can start without a complete detailed design of the whole system. | Strongly Disagree | Tend to Disagree | Neither Agree nor Disagree | Tend to Agree | Strongly Agree |
| OR2_D23 | Design can start without all the requirements being known except those that are architectural influential. | Strongly Disagree | Tend to Disagree | Neither Agree nor Disagree | Tend to Agree | Strongly Agree |
| OR2_D24 | Design should be revisited before the start of each iteration. | Strongly Disagree | Tend to Disagree | Neither Agree nor Disagree | Tend to Agree | Strongly Agree |

# *3.2.1.2. Questions to be answered by Managers/Executives*

To what extent do you agree with the statements below:

| | Statements | Nominal Values | | | | |
|---|---|---|---|---|---|---|
| | | V | W | X | Y | Z |
| OR2_M1 | The organization employees know the procedures to gather requirements from clients. | Strongly Disagree | Tend to Disagree | Neither Agree nor Disagree | Tend to Agree | Strongly Agree |
| OR2_M2 | In any project requirements are always gathered from the customer. | Strongly Disagree | Tend to Disagree | Neither Agree nor Disagree | Tend to Agree | Strongly Agree |
| OR2_M3 | Indicate how often you manage a project in which all the requirements are not known upfront and an evolutionary requirements approach is used. | Never | Seldom | Sometimes | Usually | Always |
| OR2_M4 | You can start a development of a project without knowing the exact requirements of the whole project. | Strongly Disagree | Tend to Disagree | Neither Agree nor Disagree | Tend to Agree | Strongly Agree |
| OR2_M5 | If circumstances dictate that all the details are not available before you start a project, you do not mind the uncertainty and floating targets. | Strongly Disagree | Tend to Disagree | Neither Agree nor Disagree | Tend to Agree | Strongly Agree |
| OR2_M6 | You do not mind starting a project knowing that its requirements will evolve or change in the future. | Strongly Disagree | Tend to Disagree | Neither Agree nor Disagree | Tend to Agree | Strongly Agree |
| OR2_M7 | You can tell the difference between requirements that will the influence the architecture and design of a project and requirements that will not influence it. | Strongly Disagree | Tend to Disagree | Neither Agree nor Disagree | Tend to Agree | Strongly Agree |
| OR2_M8 | In a project, you can recognize high level requirements that most probably will not change versus low level requirements that might change. | Strongly Disagree | Tend to Disagree | Neither Agree nor Disagree | Tend to Agree | Strongly Agree |
| OR2_M9 | Throughout the project the client has full right to change the requirements in order to meet his/her business needs. | Strongly Disagree | Tend to Disagree | Neither Agree nor Disagree | Tend to Agree | Strongly Agree |
| OR2_M10 | In order to deliver valuable software to clients change should be welcomed not constrained. | Strongly Disagree | Tend to Disagree | Neither Agree nor Disagree | Tend to Agree | Strongly Agree |
| OR2_M11 | An evolutionary requirements gathering approach could work better than a big upfront approach. | Strongly Disagree | Tend to Disagree | Neither Agree nor Disagree | Tend to Agree | Strongly Agree |
| OR2_M12 | You are willing to try an evolutionary requirements gathering approach. | Strongly Disagree | Tend to Disagree | Neither Agree nor Disagree | Tend to Agree | Strongly Agree |
| OR2_M13 | Software development in this organization is not ad hoc or haphazard; there is a clear and known process in place. | Strongly Disagree | Tend to Disagree | Neither Agree nor Disagree | Tend to Agree | Strongly Agree |
| OR2_M14 | Every project involves a clear set of activities. Each of these activities has clear standardized deliverables. | Strongly Disagree | Tend to Disagree | Neither Agree nor Disagree | Tend to Agree | Strongly Agree |
| OR2_M15 | Indicate how often you develop a project using an incremental–iterative approach. | Never | Seldom | Sometimes | Usually | Always |
| OR2_M16 | It is a common practice to divide the system up into mini-projects. The system is seldom developed as one large project. | Strongly Disagree | Tend to Disagree | Neither Agree nor Disagree | Tend to Agree | Strongly Agree |
| OR2_M17 | The incremental-iterative approach has more benefits than the waterfall approach. | Strongly Disagree | Tend to Disagree | Neither Agree nor Disagree | Tend to Agree | Strongly Agree |
| OR2_M18 | You are willing to use the incremental-iterative approach to develop software. | Strongly Disagree | Tend to Disagree | Neither Agree nor Disagree | Tend to Agree | Strongly Agree |
| OR2_M19 | Delivering a working increment every 1-4 weeks will not cause you any additional stress. | Strongly Disagree | Tend to Disagree | Neither Agree nor Disagree | Tend to Agree | Strongly Agree |
| OR2_M20 | No big upfront analysis should be conducted when using the incremental-iterative approach. | Strongly Disagree | Tend to Disagree | Neither Agree nor Disagree | Tend to Agree | Strongly Agree |
| OR2_M21 | You fully understand the principles of the incremental-iterative development approach. | Strongly Disagree | Tend to Disagree | Neither Agree nor Disagree | Tend to Agree | Strongly Agree |
| OR2_M22 | Planning the project from multiple levels or perspectives (iterations, releases…etc) is better than having one plan for the whole project. | Strongly Disagree | Tend to Disagree | Neither Agree nor Disagree | Tend to Agree | Strongly Agree |
| OR2_M23 | You understand the importance of planning the project in terms of iterations and releases. | Strongly Disagree | Tend to Disagree | Neither Agree nor Disagree | Tend to Agree | Strongly Agree |
| OR2_M24 | You can differentiate between planning features and planning tasks. | Strongly Disagree | Tend to Disagree | Neither Agree nor Disagree | Tend to Agree | Strongly Agree |
| OR2_M25 | Planning for each iteration should occur only right before the actual iteration. | Strongly Disagree | Tend to Disagree | Neither Agree nor Disagree | Tend to Agree | Strongly Agree |
| OR2_M26 | Planning of releases should not be detailed, except for the next/current release. | Strongly Disagree | Tend to Disagree | Neither Agree nor Disagree | Tend to Agree | Strongly Agree |
| OR2_M27 | Indicate your willingness to start a project that is not completely planned out until the end. | Strongly Disagree | Tend to Disagree | Neither Agree nor Disagree | Tend to Agree | Strongly Agree |
| OR2_M28 | Indicate your willingness to commit to planning small iteration and releases continuously through out the project and not to develop one big plan at the beginning of the project. | Strongly Disagree | Tend to Disagree | Neither Agree nor Disagree | Tend to Agree | Strongly Agree |
| OR2_M29 | Indicate how often you create multi-level planning documents when planning a project. | Never | Seldom | Sometimes | Usually | Always |
| OR2_M30 | The organization has a usable and efficient method for reporting project status. | Strongly Disagree | Tend to Disagree | Neither Agree nor Disagree | Tend to Agree | Strongly Agree |
| OR2_M31 | During development you deliver a software iteration/release at least once within the organizational status-reporting window (usually one month). | Strongly Disagree | Tend to Disagree | Neither Agree nor Disagree | Tend to Agree | Strongly Agree |
| OR2_M32 | Working software should be the primary measure of progress in a project. | Strongly Disagree | Tend to Disagree | Neither Agree nor Disagree | Tend to Agree | Strongly Agree |
| OR2_M33 | Development of the first iteration can start without a complete detailed design of the whole system. | Strongly Disagree | Tend to Disagree | Neither Agree nor Disagree | Tend to Agree | Strongly Agree |
| OR2_M34 | Design can start without all the requirements being know, except those that are architectural influential. | Strongly Disagree | Tend to Disagree | Neither Agree nor Disagree | Tend to Agree | Strongly Agree |
| OR2_M35 | Design should be revisited before the start of each iteration. | Strongly Disagree | Tend to Disagree | Neither Agree nor Disagree | Tend to Agree | Strongly Agree |
| OR2_M36 | In the organization, design is a continuous process that spans the whole development effort and is not done only one time up front. | Strongly Disagree | Tend to Disagree | Neither Agree nor Disagree | Tend to Agree | Strongly Agree |
| OR2_M37 | Indicate how often the organization does not undertake design as a big upfront activity, and instead designs in small increments throughout the development process. | Never | Seldom | Sometimes | Usually | Always |

### *3.2.1.3.Questions to be answered by the Assessor through observations*

To what extent do you agree with the statements below:

| | **Statements** | Nominal Values | | | | |
|---|---|---|---|---|---|---|
| | | V | W | X | Y | Z |
| OR2_A1 | A review of policies and procedures shows that the organization has a process it uses to gather requirements from its clients. | Strongly Disagree | Tend to Disagree | Neither Agree nor Disagree | Tend to Agree | Strongly Agree |
| OR2_A2 | A review of the policies and procedures shows that the organization has a process it uses to develop software. This process should include a set of activities with deliverables and standards. | Strongly Disagree | Tend to Disagree | Neither Agree nor Disagree | Tend to Agree | Strongly Agree |
| OR2_A3 | Inspection of the software development environment shows that the organization has sufficient and useable Software Configuration Tools for agile development. | Strongly Disagree | Tend to Disagree | Neither Agree nor Disagree | Tend to Agree | Strongly Agree |

## *3.3. Assessment Tables for Practices and Concepts in Agile Level 3*

## Risk Driven Iterations [14]

| Category of Assessment | Area to be assessed | Characteristic(s) to be assessed | To determine: | Assessment Method | Sample Indicators |
|---|---|---|---|---|---|
| People | Managers | Competence | Whether or not the managers are competent risk assessors | Interviewing | OR3_M1, OR3_M2, OR3_M3, OR3_D2 |
| | | Buy-In | Whether or not managers agree to have risks drive the scope of each iteration | Interviewing | OR3_M4 |
| | Developers | Buy-In | Whether or not the developers agree to have risks drive the scope of each iteration | Interviewing | OR3_D3 |
| Process | Risk Assessment | Experience | Whether or not the organization has any experience doing risk assessment or not | Interviewing | OR3_M1, OR3_D1 |

## Continuous Improvement [13, 4]

| Category of Assessment | Area to be assessed | Characteristic(s) to be assessed | To determine: | Assessment Method | Sample Indicators |
|---|---|---|---|---|---|
| People | Developers | Buy-in | Whether or not the developers agree to adopt an approach of continuous improvement while developing software | Interviewing | OR3_D4, OR3_D5 |
| | | Competence | Whether or not the developers are competent enough to refractor code without jeopardizing the existing functionality and quality of the code | Interviewing | OR3_M5 |
| Process | Continuous Improvement | Experience | Whether or not the organization is already involved in continuous improvement | Interviewing | OR3_D6, OR3_D7, OR3_M6, OR3_M7 |

## Self Organizing Teams

| Category of Assessment | Area to be assessed | Characteristic(s) to be assessed | To determine: | Assessment Method | Sample Indicators |
|---|---|---|---|---|---|
| People | Management | Buy-in | Whether or not management agrees to have self-organizing teams | Interviewing | OR3_M11 |
| | | Competence | Whether or not management is ready to treat the team as a true self-organizing team | Interviewing | OR3_M8, OR3_M10, OR3_M9, OR3_M12, OR3_M13 |
| | Developers | Buy-In | Whether or not the employees feel comfortable working as self-organizing teams | Interviewing | OR3_D8, OR3_D9, OR3_D10 |

## The use of True Object Oriented (OO) Design and Construction

| Category of Assessment | Area to be assessed | Characteristic(s) to be assessed | To determine: | Assessment Method | Sample Indicators |
|---|---|---|---|---|---|
| People | Developers | Competence | Whether or not the developers are experienced with object oriented design and development | Interviewing | OR3_D11, OR3_D12 |
| Process | Development | Experience | Whether or not the organization has a lot of previous experience with OO development | Interviewing | OR3_M14, OR3_D13 |

## Continuous Integration [13]

| Category of Assessment | Area to be assessed | Characteristic(s) to be assessed | To determine: | Assessment Method | Sample Indicators |
|---|---|---|---|---|---|
| People | Developers | Buy-In | Whether or not the developers are willing to commit to continuous integration? | Interviewing | OR3_D14, OR3_D15, OR3_D16, OR3_D17 |
| Environment | Software Tools | Existence | Whether or not the organization has the tools to aid in continuous integration | Observation | OR3_A1 |

## Maintenance of a List of All Remaining Features (Backlog)

| Category of Assessment | Area to be assessed | Characteristic(s) to be assessed | To determine: | Assessment Method | Sample Indicators |
|---|---|---|---|---|---|
| People | Management | Buy-in | Whether or not management is willing to maintain an up-to-date list of all the remaining features for the project (backlog) | Interviewing | OR3_M16 |
| Process | Project Management | Existence | Whether or not the organization currently keeps an up-to-date list of all the work that remains to be done | Interviewing | OR3_M15 |
| | | | | Observation | OR3_A2 |

## Unit Tests

| Category of Assessment | Area to be assessed | Characteristic(s) to be assessed | To determine: | Assessment Method | Sample Indicators |
|---|---|---|---|---|---|
| People | Developers | Buy-in | Whether or not developers are willing to write unit tests during the development process | Interviewing | OR3_D18, OR3_D19, OR3_D21 |
| | | Competence | Whether or not the developers have the competence and previous experience writing unit tests | Interviewing | OR3_D20, OR3_M19, OR3_D22 |
| | Managers | Buy-In | Whether or not the management accepts that developers will invest additional time to write unit tests while coding | Interviewing | OR3_M17, OR3_M18 |
| Environment | Software Tools | Existence | Whether or not the organization has the tools that support writing and running unit tests | Observation | OR3_A3 |

# 3.3.1. Indicators for Agile Level 3

## 3.3.1.1. Questions to be answered by Developers

To what extent do you agree with the statements below:

| | Statements | Nominal Values | | | | |
|---|---|---|---|---|---|---|
| | | V | W | X | Y | Z |
| OR3_D1 | For the projects that you have worked on, indicate how often risk assessment was performed during the project and communicated to the whole team. | Never | Seldom | Sometimes | Usually | Always |
| OR3_D2 | Your manager is very competent when coming to risk assessments and mitigation plans. | Strongly Disagree | Tend to Disagree | Neither Agree nor Disagree | Tend to Agree | Strongly Agree |
| OR3_D3 | The riskiest, most difficult elements should be approached first in the early iterations of the development effort. | Strongly Disagree | Tend to Disagree | Neither Agree nor Disagree | Tend to Agree | Strongly Agree |
| OR3_D4 | It is important to put effort into improving the design and code of a component, even if it is already working. | Strongly Disagree | Tend to Disagree | Neither Agree nor Disagree | Tend to Agree | Strongly Agree |
| OR3_D5 | You are willing to adopt an approach of continuous improvement during software development. | Strongly Disagree | Tend to Disagree | Neither Agree nor Disagree | Tend to Agree | Strongly Agree |
| OR3_D6 | It is a common practice in the organization to revisit a working component to improve its design or code structure. | Strongly Disagree | Tend to Disagree | Neither Agree nor Disagree | Tend to Agree | Strongly Agree |
| OR3_D7 | Indicate how often you revisit a working component to improve its design or code structure. | Never | Seldom | Sometimes | Usually | Always |
| OR3_D8 | You like to work on a team that management regards as one entity; not addressing individual team members in rewards or tasks, but as one team. | Strongly Disagree | Tend to Disagree | Neither Agree nor Disagree | Tend to Agree | Strongly Agree |
| OR3_D9 | You do not mind working without direct managerial supervision as long as you are on a team that is treated as a partner with management. | Strongly Disagree | Tend to Disagree | Neither Agree nor Disagree | Tend to Agree | Strongly Agree |
| OR3_D10 | You consider yourself competent and disciplined enough to work on self-organizing teams | Strongly Disagree | Tend to Disagree | Neither Agree nor Disagree | Tend to Agree | Strongly Agree |
| OR3_D11 | Indicate how often you develop software projects using the Object Oriented (OO) principles and techniques. | Never | Seldom | Sometimes | Usually | Always |
| OR3_D12 | You understand the OO principles and theories very well. | Strongly Disagree | Tend to Disagree | Neither Agree nor Disagree | Tend to Agree | Strongly Agree |
| OR3_D13 | Indicate how often the organization takes the OO approach in development of software projects. | Strongly Disagree | Tend to Disagree | Neither Agree nor Disagree | Tend to Agree | Strongly Agree |
| OR3_D14 | The usual time it takes to create a build the system is: | More than 1 hour | Under 1 hour | Under 15 minutes | Under 10 minutes | Under 5 minutes |
| OR3_D15 | Instead of integrating the system at the end of the development effort, it is better to regularly integrate the system throughout the whole development process. | Strongly Disagree | Tend to Disagree | Neither Agree nor Disagree | Tend to Agree | Strongly Agree |
| OR3_D16 | You are trained to use the Software Configuration Management tool for continuous integration. | Strongly Disagree | Tend to Disagree | Neither Agree nor Disagree | Tend to Agree | Strongly Agree |
| OR3_D17 | You are willing to integrate your software throughout the development process, even if it means more work for you. | Strongly Disagree | Tend to Disagree | Neither Agree nor Disagree | Tend to Agree | Strongly Agree |
| OR3_D18 | It is important to write unit tests for methods and functions while coding them even if that will take additional time. | Strongly Disagree | Tend to Disagree | Neither Agree nor Disagree | Tend to Agree | Strongly Agree |
| OR3_D19 | Writing unit tests for code is as important as writing new code for more functionality. | Strongly Disagree | Tend to Disagree | Neither Agree nor Disagree | Tend to Agree | Strongly Agree |
| OR3_D20 | Indicate how often you write unit tests for every method or function in your code. | Never | Seldom | Sometimes | Usually | Always |
| OR3_D21 | You are willing to commit to writing unit tests while you code for every method or function in your code. | Strongly Disagree | Tend to Disagree | Neither Agree nor Disagree | Tend to Agree | Strongly Agree |
| OR3_D22 | You consider yourself competent enough to write good and comprehensive unit tests for the methods and functions in your code. | Strongly Disagree | Tend to Disagree | Neither Agree nor Disagree | Tend to Agree | Strongly Agree |

## *3.3.1.2.Questions to be answered by Managers/Executives*

To what extent do you agree with the statements below:

| | Statements | Nominal Values | | | | |
|---|---|---|---|---|---|---|
| | | V | W | X | Y | Z |
| OR3_M1 | Indicate how often do you perform risk assessment and mitigation techniques during a project. | Never | Seldom | Sometimes | Usually | Always |
| OR3_M2 | You have been trained to perform risk assessments. | Strongly Disagree | Tend to Disagree | Neither Agree nor Disagree | Tend to Agree | Strongly Agree |
| OR3_M3 | You are very competent performing risk assessment and mitigation plans. | Strongly Disagree | Tend to Disagree | Neither Agree nor Disagree | Tend to Agree | Strongly Agree |
| OR3_M4 | The riskiest, most difficult elements should be approached first in the early iterations of the development effort. | Strongly Disagree | Tend to Disagree | Neither Agree nor Disagree | Tend to Agree | Strongly Agree |
| OR3_M5 | The developers are competent enough to refractor code without jeopardizing the existing functionality and quality and breaking any unit tests (if they exist). | Strongly Disagree | Tend to Disagree | Neither Agree nor Disagree | Tend to Agree | Strongly Agree |
| OR3_M6 | It is a common practice in the organization to revisit a working component to improve its design or code structure. | Strongly Disagree | Tend to Disagree | Neither Agree nor Disagree | Tend to Agree | Strongly Agree |
| OR3_M7 | Indicate how often you make sure that your subordinates revisit a working component to improve its design or code structure. | Never | Seldom | Sometimes | Usually | Always |
| OR3_M8 | You can trust your employees' capabilities to determine the best way to accomplish tasks by themselves without your (management's) interference. | Strongly Disagree | Tend to Disagree | Neither Agree nor Disagree | Tend to Agree | Strongly Agree |
| OR3_M9 | Employees are competent and disciplined enough to work in self-organizing teams. | Strongly Disagree | Tend to Disagree | Neither Agree nor Disagree | Tend to Agree | Strongly Agree |
| OR3_M10 | You are willing to allow space for the self-organizing team to grow and not micromanage it. | Strongly Disagree | Tend to Disagree | Neither Agree nor Disagree | Tend to Agree | Strongly Agree |
| OR3_M11 | You agree that it is very important for the employees to work in teams where they can divide the team tasks among themselves. | Strongly Disagree | Tend to Disagree | Neither Agree nor Disagree | Tend to Agree | Strongly Agree |
| OR3_M12 | The team is an entity that has its knowledge, perspective, motivation and expertise and should be treated as a partner with management and the customer. | Strongly Disagree | Tend to Disagree | Neither Agree nor Disagree | Tend to Agree | Strongly Agree |
| OR3_M13 | The self-organizing team can negotiate commitments. | Strongly Disagree | Tend to Disagree | Neither Agree nor Disagree | Tend to Agree | Strongly Agree |
| OR3_M14 | Indicate how often your organization takes the OO approach in software development | Never | Seldom | Sometimes | Usually | Always |
| OR3_M15 | When working on a project you keep an up-to-date list of all the work that remains to be done. | Strongly Disagree | Tend to Disagree | Neither Agree nor Disagree | Tend to Agree | Strongly Agree |
| OR3_M16 | You are willing to keep an up-to-date list of all the work that remains to be done. | Strongly Disagree | Tend to Disagree | Neither Agree nor Disagree | Tend to Agree | Strongly Agree |
| OR3_M17 | It is important for developers to write unit tests for their methods and functions while they code, even if that will take additional time from them. | Strongly Disagree | Tend to Disagree | Neither Agree nor Disagree | Tend to Agree | Strongly Agree |
| OR3_M18 | Writing unit tests for code is as important as writing new code for more functionality. | Strongly Disagree | Tend to Disagree | Neither Agree nor Disagree | Tend to Agree | Strongly Agree |
| OR3_M19 | The developers are competent enough to write good unit tests for the methods and functions in the code. | Strongly Disagree | Tend to Disagree | Neither Agree nor Disagree | Tend to Agree | Strongly Agree |

## *3.3.1.3. Questions to be answered by the Assessor through observations*

To what extent do you agree with the statements below:

| | **Statements** | Nominal Values | | | | |
|---|---|---|---|---|---|---|
| | | V | W | X | Y | Z |
| OR3_A1 | After looking at the software development tools, you know that the organization has the SCM tools and processes to support continuous integration. | Strongly Disagree | Tend to Disagree | Neither Agree nor Disagree | Tend to Agree | Strongly Agree |
| OR3_A2 | After inspecting previous projects, you know that each project had a mechanism by which all the remaining work in a project was known at any point in time. | Strongly Disagree | Tend to Disagree | Neither Agree nor Disagree | Tend to Agree | Strongly Agree |
| OR3_A3 | After looking at the software development tools, you know that the organization has the necessary tools to write and run unit tests within the development IDE. | Strongly Disagree | Tend to Disagree | Neither Agree nor Disagree | Tend to Agree | Strongly Agree |

## *3.4. Assessment Tables for Practices and Concepts in Agile Level 4*

## Client Driven Iterations

| Category of Assessment | Area to be assessed | Characteristic(s) to be assessed | To determine: | Assessment Method | Sample Indicators |
|---|---|---|---|---|---|
| People | Managers | Buy-in | Whether or not managers are willing to give the customer the power to dictate the scope of the iterations | Interviewing | OR4_M1, OR4_M2, OR4_M3 |

## Continuous Customer Satisfaction Feedback

| Category of Assessment | Area to be assessed | Characteristic(s) to be assessed | To determine: | Assessment Method | Sample Indicators |
|---|---|---|---|---|---|
| Process | Customer Feedback | Existence | Whether or not the organization has a method by which they gather continuous feedback/criticism from the customer during the development process | Interviewing | OR4_M4, OR4_M5 |
| People | Developers | Buy-in | Whether or not the developers accept the fact that the customers are encouraged to continually re-think their requirements | Interviewing | OR4_D1, OR4_D2, OR4_D3 |
| | Managers | Buy-in | Whether or not the managers accept the fact that the customers are encouraged to continually re-think their requirements | Interviewing | OR4_M2, OR4_M6, OR4_M7 |

## Smaller and more Frequent Releases

| Category of Assessment | Area to be assessed | Characteristic(s) to be assessed | To determine: | Assessment Method | Sample Indicators |
|---|---|---|---|---|---|
| People | Managers | Buy-in | Whether or not the managers understand the importance of having smaller and more frequent releases to give the customer quicker feedback | Interviewing | OR4_M12 |
| | | Stress | Whether or not managers can handle the additional stress of overseeing the delivery of fully functional releases every 4-8 weeks | Interviewing | OR4_M13 |
| | Developers | Buy-in | Whether or not the developers understand the importance of having smaller and more frequent releases to give the customer quicker feedback | Interviewing | OR4_D8 |
| | | Stress | Whether or not the developers can handle the increased stress of delivering fully functional releases every 4-8 weeks | Interviewing | OR4_D9 |

## Adaptive Planning

| Category of Assessment | Area to be assessed | Characteristic(s) to be assessed | To determine: | Assessment Method | Sample Indicators |
|---|---|---|---|---|---|
| People | Management | Buy-in | Whether or not management is willing to base the planning for the next iteration on the client's feedback from the current (previous) iteration | Interviewing | OR4_M14 |
| | | | Whether or not management is willing to plan as late as possible for an iteration (immediately before the iteration) | Interviewing | OR4_M15 |

## Daily Progress Tracking Meetings

| Category of Assessment | Area to be assessed | Characteristic(s) to be assessed | To determine: | Assessment Method | Sample Indicators |
|---|---|---|---|---|---|
| People | Management | Buy-In | Whether or not management is willing to meet daily for progress update | Interviewing | OR4_M16 |
| | Developers | Buy-In | Whether or not the developers are willing to meet daily for progress updates | Interviewing | OR4_D10 |
| Process | Project management | Progress meetings | How often the team meets regularly to discuss the progress of a project | Interviewing | OR4_M17, OR4_D11 |

## Agile Documentation (from Agile Modeling)

| Category of Assessment | Area to be assessed | Characteristic(s) to be assessed | To determine: | Assessment Method | Sample Indicators |
|---|---|---|---|---|---|
| People | Developers | Competence | Whether or not the developers understand what an Agile approach to documentation is | Interviewing | OR4_D12, OR4_D13 |
| | Management | Competence | Whether or not management understands what an Agile approach to documentation is | Interviewing | OR4_M18, OR4_M19 |
| | | Buy-In | Whether or not management is willing to take an Agile approach to documentation | Interviewing | OR4_M20 |
| Process | Documentation | Regulations | Whether or not external regulatory requirements exist that dictate the production of heavy (detailed) documentation for every aspect of the process | Interviewing | OR4_M21, OR4_M22 |

## User Stories

| Category of Assessment | Area to be assessed | Characteristic(s) to be assessed | To determine: | Assessment Method | Sample Indicators |
|---|---|---|---|---|---|
| People | Management | Buy-In | Whether or not management is willing to use user stories as an elicitation method/form for high level requirements | Interviewing | OR4_M23, OR4_M24 |
| | Developers | Competence | Whether or not the developers have the understanding/knowledge of how to use user stories | Interviewing | OR4_D14 |
| Process | Requirements | Regulations | Whether or not there are regulatory requirements for the elicitation of the requirements (they have to specified in a certain form) | Interviewing | OR4_M25 |

## *3.4.1.Indicators for Agile Level 4*

### *3.4.1.1.Questions to be answered by Developers*

To what extent do you agree with the statements below:

| | Statements | Nominal Values | | | | |
|---|---|---|---|---|---|---|
| | | V | W | X | Y | Z |
| OR4_D1 | Customers should be encouraged to regularly change their expectations for the product being developed to ensure that the product satisfies the business priorities of the organization. | Strongly Disagree | Tend to Disagree | Neither Agree nor Disagree | Tend to Agree | Strongly Agree |
| OR4_D2 | As the perception of what they need changes, customers are expected to articulate these changes and thus affect the product being built. | Strongly Disagree | Tend to Disagree | Neither Agree nor Disagree | Tend to Agree | Strongly Agree |
| OR4_D3 | The customer should give his/her feedback throughout the development process even if it means that requirements must be changed. | Strongly Disagree | Tend to Disagree | Neither Agree nor Disagree | Tend to Agree | Strongly Agree |
| OR4_D8 | Smaller and more frequent releases are important in order to give the customer a means by which he/she can give more and quicker feedback. | Strongly Disagree | Tend to Disagree | Neither Agree nor Disagree | Tend to Agree | Strongly Agree |
| OR4_D9 | Delivering smaller and more frequent fully functional releases every 4-8 weeks will not cause you any additional stress. | Strongly Disagree | Tend to Disagree | Neither Agree nor Disagree | Tend to Agree | Strongly Agree |
| OR4_D10 | You are willing to meet daily for the progress update of a project. | Strongly Disagree | Tend to Disagree | Neither Agree nor Disagree | Tend to Agree | Strongly Agree |
| OR4_D11 | Indicate how often you meet with the rest of the team to discuss and update each other about the progress of the project. | Less frequent than monthly | Monthly | Every couple of weeks | Weekly | Daily/Hourly |
| OR4_D12 | Documentation exists within an Agile development process | Strongly Disagree | Tend to Disagree | Neither Agree nor Disagree | Tend to Agree | Strongly Agree |
| OR4_D13 | You understand the role of documentation within an Agile development process | Strongly Disagree | Tend to Disagree | Neither Agree nor Disagree | Tend to Agree | Strongly Agree |
| OR4_D14 | You can use user stories instead of requirements to develop the architectural framework of the system. | Strongly Disagree | Tend to Disagree | Neither Agree nor Disagree | Tend to Agree | Strongly Agree |

## *3.4.1.2.Questions to be answered by Managers/Executives*

To what extent do you agree with the statements below:

| | **Statements** | Nominal Values | | | | |
|---|---|---|---|---|---|---|
| | | V | W | X | Y | Z |
| OR4_M1 | As the perception of what they need changes, customers are expected to articulate those changes by prioritizing the features they would like to see in the next iteration. | Strongly Disagree | Tend to Disagree | Neither Agree nor Disagree | Tend to Agree | Strongly Agree |
| OR4_M2 | Customers should be encouraged to regularly change their expectations for the product being developed to ensure that the product satisfies the business priorities of the organization. | Strongly Disagree | Tend to Disagree | Neither Agree nor Disagree | Tend to Agree | Strongly Agree |
| OR4_M3 | The customer should be given the authority to direct what is being developed in which iteration. | Strongly Disagree | Tend to Disagree | Neither Agree nor Disagree | Tend to Agree | Strongly Agree |
| OR4_M4 | The customer has the opportunity to give his/her feedback about the product through out the development process by means of interacting with a working piece of software or a least a prototype. | Strongly Disagree | Tend to Disagree | Neither Agree nor Disagree | Tend to Agree | Strongly Agree |
| OR4_M5 | The organization has a method by which it gathers continuous feedback/criticism from the customer during the development process. | Strongly Disagree | Tend to Disagree | Neither Agree nor Disagree | Tend to Agree | Strongly Agree |
| OR4_M6 | As the perception of what they need changes, customers are expected to articulate those changes and so affect the product being built. | Strongly Disagree | Tend to Disagree | Neither Agree nor Disagree | Tend to Agree | Strongly Agree |
| OR4_M7 | The customer should give his/her feedback throughout the development process even if it means that requirements must be changed. | Strongly Disagree | Tend to Disagree | Neither Agree nor Disagree | Tend to Agree | Strongly Agree |
| OR4_M12 | Smaller and more frequent releases are important in order to give the customer a means by which he/she can give more and quicker feedback. | Strongly Disagree | Tend to Disagree | Neither Agree nor Disagree | Tend to Agree | Strongly Agree |
| OR4_M13 | Delivering smaller and more frequent fully functional releases every 4-8 weeks will not cause you any additional stress. | Strongly Disagree | Tend to Disagree | Neither Agree nor Disagree | Tend to Agree | Strongly Agree |
| OR4_M14 | The plan for upcoming iteration may change based on customer feedback from the previous or current iteration. | Strongly Disagree | Tend to Disagree | Neither Agree nor Disagree | Tend to Agree | Strongly Agree |
| OR4_M15 | You agree with developing the detailed plan for an iteration only after the conclusion of the previous iteration. | Strongly Disagree | Tend to Disagree | Neither Agree nor Disagree | Tend to Agree | Strongly Agree |
| OR4_M16 | You are willing to meet daily for the progress update of a project. | Strongly Disagree | Tend to Disagree | Neither Agree nor Disagree | Tend to Agree | Strongly Agree |
| OR4_M17 | Indicate how often you meet with the rest of the team to discuss and update each other on the progress of the project. | Less frequent | Monthly | Every couple of weeks | Weekly | Daily/Hourly |
| | | than monthly | | | | |
| OR4_M18 | Documentation exists within an Agile development process. | Strongly Disagree | Tend to Disagree | Neither Agree nor Disagree | Tend to Agree | Strongly Agree |
| OR4_M19 | You understand the role of documentation within an Agile development process. | Strongly Disagree | Tend to Disagree | Neither Agree nor Disagree | Tend to Agree | Strongly Agree |
| OR4_M20 | You will allow your subordinates to take an Agile approach to documentation. | Strongly Disagree | Tend to Disagree | Neither Agree nor Disagree | Tend to Agree | Strongly Agree |
| OR4_M21 | Stakeholders do not require heavy (detailed) documentation for any activities or aspects of the development process. | Strongly Disagree | Tend to Disagree | Neither Agree nor Disagree | Tend to Agree | Strongly Agree |
| OR4_M22 | You are not required by any external auditory to maintain fine heavy (detailed) documentation for activities or aspects of the development process. | Strongly Disagree | Tend to Disagree | Neither Agree nor Disagree | Tend to Agree | Strongly Agree |
| OR4_M23 | You are willing to adopt user stories as a method for high level requirements elicitation. | Strongly Disagree | Tend to Disagree | Neither Agree nor Disagree | Tend to Agree | Strongly Agree |
| OR4_M24 | User stories can be used instead of large requirements documents. | Strongly Disagree | Tend to Disagree | Neither Agree nor Disagree | Tend to Agree | Strongly Agree |
| OR4_M25 | No regulatory constraints exist that prevent the use of user stories as a means of capturing high level requirements from the user. | Strongly Disagree | Tend to Disagree | Neither Agree nor Disagree | Tend to Agree | Strongly Agree |

## *3.5. Assessment Tables for Practices and Concepts in Agile Level 5*

## **Low Process Ceremony** (Process Ceremony is the level of paperwork involved with a process)

| Category of Assessment | Area to be assessed | Characteristic(s) to be assessed | To determine: | Assessment Method | Sample Indicators |
|---|---|---|---|---|---|
| Process | Ceremony | Regulations | Whether or not the organization needs to maintain a high process ceremony due to certain audits or regulations | Interviewing | OR5_M1, OR5_M2 |
| | | | | Observation | OR5_A1 |
| Culture | Organizational | Responsibility | Whether or not there is a fear of responsibility/blame among people, thus supporting the high level of process ceremony | Interviewing | OR5_M3 |
| People | Developers | Buy-in | Whether or not the developers feel comfortable decreasing the level of process ceremony | Interviewing | OR5_D1, OR5_D2 |
| | Management | Buy-in | Whether or not the managers feel comfortable decreasing the level of process ceremony | Interviewing | OR5_M4, OR5_M5 |
| | | Trust | Whether or not the management trusts the developers to make decisions on their own without their "approval" | Interviewing | OR5_M6, OR5_M7 |

## Agile Project Estimation

| Category of Assessment | Area to be assessed | Characteristic(s) to be assessed | To determine: | Assessment Method | Sample Indicators |
|---|---|---|---|---|---|
| Process | Estimation | Experience | Whether or not the organization is experienced in estimation | Observation | OR5_A2 |
| | | Existence | Whether or not data exists from previous projects to aid with the estimation | Observation | OR5_A3 |
| | | Method | Whether or not the estimation process separates the estimation of effort from the estimation of duration | Interviewing | OR5_M8, OR5_M9 |
| People | Developers | Competence | Whether or not the developers are competent in making their own estimates of effort. | Interviewing | OR5_D3, OR5_D4, OR5_D5 |
| | Management | Competence | Whether or not the managers are competent in making estimates. | Interviewing | OR5_M10, OR5_M11, OR5_M12 |
| | Management | Collaboration | Whether management will encourage the estimation process to be done by the whole team or by only them | Interviewing | OR5_M13, OR5_M14, OR5_M15 |

## Paired Programming

| Category of Assessment | Area to be assessed | Characteristic(s) to be assessed | To determine: | Assessment Method | Sample Indicators |
|---|---|---|---|---|---|
| People | Management | Buy-in | Whether or not management can see the benefit from paired programming | Interviewing | OR5_M16, OR5_M17 |
| | Developers | Buy-in | Whether or not developers are willing to try paired programming | Interviewing | OR5_D6, OR5_D7, OR5_D8 |
| Process | Project Management | Measurement of Productivity | What the organization considers to be a measure of software productivity | Interviewing | OR5_M18 |
| Culture | Organizational | Collaboration | Whether or not an atmosphere of assistance exists in the organization | Interviewing | OR5_D9, OR5_D10, OR5_M19 |

## Test Driven Development

| Category of Assessment | Area to be assessed | Characteristic(s) to be assessed | To determine: | Assessment Method | Sample Indicators |
|---|---|---|---|---|---|
| People | Developers | Competence | Whether or not the developers are competent and experienced with writing unit tests | Interviewing | OR5_D11, OR5_D12, OR5_D13 |
| | | | Whether or not the developers have a very strong understanding of OO concepts | Interviewing | OR5_D14, OR5_D15, OR5_M20 |
| | | Buy-In | Whether or not the developers are motivated and willing to apply test driven development | Interviewing | OR5_D16 |
| | | Perception | Whether or not the developers think that Test-driven development is a hard task or not | Interviewing | OR5_D17 |
| | Management | Buy-In | Whether or not management will encourage test-driven development and tolerate the learning curve | Interviewing | OR5_M21. OR5_M22 |
| Environment | Software Tools | Test Automation | Whether or not the organization has or can provide tools for creating and maintaining automated test suites | Observation | OR5_A4 |
| | | | | Interviewing | OR5_M23 |

# 3.5.1.Indicators for Agile Level 5

## 3.5.1.1.Questions to be answered by Developers

To what extent do you agree with the statements below:

| | Statements | Nominal Values | | | | |
|---|---|---|---|---|---|---|
| | | V | W | X | Y | Z |
| OR5_D1 | You favor accepting responsibility and being held accountable when things go wrong over multiple layers of formal steps, reviews, and procedures. | Strongly Disagree | Tend to Disagree | Neither Agree nor Disagree | Tend to Agree | Strongly Agree |
| OR5_D2 | You do not support the existence of various formal steps and reviews to reduce (spread) accountability when something goes wrong. | Strongly Disagree | Tend to Disagree | Neither Agree nor Disagree | Tend to Agree | Strongly Agree |
| OR5_D3 | Indicate how often you make size/effort estimates for the project or a component of the project that you will be working on. | Never | Seldom | Sometimes | Usually | Always |
| OR5_D4 | You have been trained on how to make project estimates. | Strongly Disagree | Tend to Disagree | Neither Agree nor Disagree | Tend to Agree | Strongly Agree |
| OR5_D5 | You are competent enough to make your own estimates of size/effort. | Strongly Disagree | Tend to Disagree | Neither Agree nor Disagree | Tend to Agree | Strongly Agree |
| OR5_D6 | Paired programming increases productivity contrary to what others say about paired programming decreasing productivity by half. | Strongly Disagree | Tend to Disagree | Neither Agree nor Disagree | Tend to Agree | Strongly Agree |
| OR5_D7 | Indicate how often you program in pairs. | Never | Seldom | Sometimes | Usually | Always |
| OR5_D8 | You are willing to program in pairs. | Strongly Disagree | Tend to Disagree | Neither Agree nor Disagree | Tend to Agree | Strongly Agree |
| OR5_D9 | An atmosphere of assistance exists in the organization. | Strongly Disagree | Tend to Disagree | Neither Agree nor Disagree | Tend to Agree | Strongly Agree |
| OR5_D10 | Whenever you need help people are willing to help you. | Strongly Disagree | Tend to Disagree | Neither Agree nor Disagree | Tend to Agree | Strongly Agree |
| OR5_D11 | Indicate how often you write unit tests for every function or method one writing code. | Never | Seldom | Sometimes | Usually | Always |
| OR5_D12 | You have no problems or challenges writing unit tests for functions and methods. | Strongly Disagree | Tend to Disagree | Neither Agree nor Disagree | Tend to Agree | Strongly Agree |
| OR5_D13 | The suite of unit tests that you write is comprehensive and usually encompasses all possible test scenarios. | Strongly Disagree | Tend to Disagree | Neither Agree nor Disagree | Tend to Agree | Strongly Agree |
| OR5_D14 | Indicate how often you program in the object-oriented (OO) paradigm. | Never | Seldom | Sometimes | Usually | Always |
| OR5_D15 | You have a very strong understanding of object-oriented concepts and principles. | Strongly Disagree | Tend to Disagree | Neither Agree nor Disagree | Tend to Agree | Strongly Agree |
| OR5_D16 | You are willing to employ a test-driven approach to development. | Strongly Disagree | Tend to Disagree | Neither Agree nor Disagree | Tend to Agree | Strongly Agree |
| OR5_D17 | Test-driven development is easy. | Strongly Disagree | Tend to Disagree | Neither Agree nor Disagree | Tend to Agree | Strongly Agree |

## *3.5.1.2. Questions to be answered by Managers/Executives*

To what extent do you agree with the statements below:

| | Statements | Nominal Values | | | | |
|---|---|---|---|---|---|---|
| | | V | W | X | Y | Z |
| OR5_M1 | There are no regulations or auditory requirements that dictate the need for high process ceremony. | Strongly Disagree | Tend to Disagree | Neither Agree nor Disagree | Tend to Agree | Strongly Agree |
| OR5_M2 | Your organization is informal and flexible. There are not many formal steps, policies or procedures. | Strongly Disagree | Tend to Disagree | Neither Agree nor Disagree | Tend to Agree | Strongly Agree |
| OR5_M3 | People in the organization are not afraid of taking responsibility. | Strongly Disagree | Tend to Disagree | Neither Agree nor Disagree | Tend to Agree | Strongly Agree |
| OR5_M4 | You favor accepting responsibility and being held accountable when things go wrong over multiple layers of formal steps, reviews, and procedures. | Strongly Disagree | Tend to Disagree | Neither Agree nor Disagree | Tend to Agree | Strongly Agree |
| OR5_M5 | You do not support the existence of various formal steps and reviews to reduce (spread) accountability when something goes wrong. | Strongly Disagree | Tend to Disagree | Neither Agree nor Disagree | Tend to Agree | Strongly Agree |
| OR5_M6 | You trust your subordinates to make decisions within their scope of work without referring back to you for approval. | Strongly Disagree | Tend to Disagree | Neither Agree nor Disagree | Tend to Agree | Strongly Agree |
| OR5_M7 | Your subordinates are competent to make their own decisions without referring back to you for approval. | Strongly Disagree | Tend to Disagree | Neither Agree nor Disagree | Tend to Agree | Strongly Agree |
| OR5_M8 | When preparing a project estimation, you estimate the size first and derive from that a duration estimate. | Strongly Disagree | Tend to Disagree | Neither Agree nor Disagree | Tend to Agree | Strongly Agree |
| OR5_M9 | The estimation process employed by the organization separates the estimation of effort from the estimation of duration. | Strongly Disagree | Tend to Disagree | Neither Agree nor Disagree | Tend to Agree | Strongly Agree |
| OR5_M10 | Indicate how often you make size/effort estimates for projects. | Never | Seldom | Sometimes | Usually | Always |
| OR5_M11 | You have been trained on how to make project estimates. | Strongly Disagree | Tend to Disagree | Neither Agree nor Disagree | Tend to Agree | Strongly Agree |
| OR5_M12 | You are competent and experienced enough to make realistic estimates of size/effort. | Strongly Disagree | Tend to Disagree | Neither Agree nor Disagree | Tend to Agree | Strongly Agree |
| OR5_M13 | The whole team participating in project estimation will yield better and more accurate results. | Strongly Disagree | Tend to Disagree | Neither Agree nor Disagree | Tend to Agree | Strongly Agree |
| OR5_M14 | Indicate how often the whole team has participated in creating project estimates. | Never | Seldom | Sometimes | Usually | Always |
| OR5_M15 | You will encourage the whole development team to actively participate in developing a project estimate. | Strongly Disagree | Tend to Disagree | Neither Agree nor Disagree | Tend to Agree | Strongly Agree |
| OR5_M16 | Paired programming increases productivity contrary to what others say about paired programming decreasing productivity by half. | Strongly Disagree | Tend to Disagree | Neither Agree nor Disagree | Tend to Agree | Strongly Agree |
| OR5_M17 | You encourage your development team to use paired programming. | Strongly Disagree | Tend to Disagree | Neither Agree nor Disagree | Tend to Agree | Strongly Agree |
| OR5_M18 | Productivity is about how much customer value can you create per dollar spent not about how many lines of code, classes coded or Use Cases per dollar spent. | Strongly Disagree | Tend to Disagree | Neither Agree nor Disagree | Tend to Agree | Strongly Agree |
| OR5_M19 | An atmosphere of assistance exists in the organization. | Strongly Disagree | Tend to Disagree | Neither Agree nor Disagree | Tend to Agree | Strongly Agree |
| OR5_M20 | The development team has a very strong understanding of object-oriented concepts and principles. | Strongly Disagree | Tend to Disagree | Neither Agree nor Disagree | Tend to Agree | Strongly Agree |
| OR5_M21 | Test-driven development will produce better software with fewer bugs | Strongly Disagree | Tend to Disagree | Neither Agree nor Disagree | Tend to Agree | Strongly Agree |
| OR5_M22 | You are willing to tolerate the learning curve of the development team while they transition to test-driven development. | Strongly Disagree | Tend to Disagree | Neither Agree nor Disagree | Tend to Agree | Strongly Agree |
| OR5_M23 | The organization will be willing to provide software tools for creating and maintaining automated test suites. | Strongly Disagree | Tend to Disagree | Neither Agree nor Disagree | Tend to Agree | Strongly Agree |

## 3.5.1.3.Questions to be answered by the Assessor through observation

To what extent do you agree with the statements below:

| | Statements | Nominal Values | | | | |
|---|---|---|---|---|---|---|
| | | V | W | X | Y | Z |
| OR5_A1 | A review of policies and procedures shows that there is no need for a high process ceremony in this organization. | Strongly Disagree | Tend to Disagree | Neither Agree nor Disagree | Tend to Agree | Strongly Agree |
| OR5_A2 | A review of previous project documentation shows that the effort estimates were within acceptable range to the actual effort that was put into delivering the project. | Strongly Disagree | Tend to Disagree | Neither Agree nor Disagree | Tend to Agree | Strongly Agree |
| OR5_A3 | Previous project documentation, including effort and size estimations, are available and accessible. | Strongly Disagree | Tend to Disagree | Neither Agree nor Disagree | Tend to Agree | Strongly Agree |
| OR5_A4 | An inspection of these software tools shows that the organization has accessible and usable tools for creating and maintaining automated test suites. | Strongly Disagree | Tend to Disagree | Neither Agree nor Disagree | Tend to Agree | Strongly Agree |

# Appendix A: The 5 Agile Levels

# The 5 Agile Levels

The process framework needs an agile measurement index to determine what degree of agility the project or organization can adopt. Both the project level assessment (Stage 2) and the organizational readiness assessment (Stage 3) need to by measured against some scale of agility. The 5 Agile levels defined in this section are considered the agile measurement index used by the process framework.

The agile levels are inspired from the core agile values and beliefs as defined by the agile manifesto not from a particular agile method. The topic of agility assessment is significant and very detailed, however since the focus of this document is credibility assessment the 5 levels of agility will be presented in a level of detail sufficient for the task of credibility assessment and without going into too much detail.

***Each agile level focuses on instilling a particular core value of agility in the software development process. The full adoption of that core value in the software process signifies that achievement of that agile level.***

For example, agile level 1 focuses on making the software development process collaborative through enhancing communication and cooperation. Agile level 4 focuses on making the software development process adaptive by responding to change through multiple levels of feedback.

| Agile level | Name | Objective |
|---|---|---|
| Level 1 | Collaborative | Enhancing communication and collaboration |
| Level 2 | Evolutionary | Delivering software early and continuously |
| Level 3 | Effective | Producing quality working software |
| Level 4 | Adaptive | Responding to change through multiple levels of feedback |
| Level 5 | Ambient | Establishing the ideal agile environment and surroundings |

**Table 4. The 5 Agile Levels defined within the process framework**

Each Agile level is defined through 2 dimensions:

1. Agile Principles
2. Agile Practices and Concepts.

## *Agile Principles*

Agile principles are guidelines or approaches that need to be employed to realize the value of each Agile Level.

For example:

- Human Centric (meaning the reliance on people and the interaction between them) is a key agile principle
- Technical Excellence (promoting the use of techniques that produce and maintain the highest quality of code possible) is another key agile principles

***Within this agile Measurement Index, all of the levels of agility are achieved by adhering to these principles. The degree and scope of adherence to the principal dictates the level of agility.***

For example, technical excellence is a key agile principle. Technical excellence is observed in agile level 1 at a very basic level and with a focus on collaboration. While at agile level 4, technical excellence is realized at a much higher degree and with the focus on being adaptive (responding to change)

The agile principles that are used by this measurement index stem from the Agile Principles that are defined directly by the Agile Manifesto. After careful analysis of the 12 principles defined by the Manifesto, we produced a condensed set of five agile principles that are used to guide the populating  the five agile levels we defined earlier. These five agile principles are:

- **Delivery customer value by embracing change** [10, 4]**:** the true success indicator for any software development effort is whether it helps to deliver customer value or not.  In many cases the development team, as well as the customer, are in a continuous learning process about the requirements necessary to realize that additional customer value.  Hence, an attitude of welcoming and embracing change should be maintained throughout the software development effort.
- **Planning to deliver software frequently** [5] [16, 9]**:** early and frequent delivery of working software is crucial because it provides the customer with functionality which they can review and give feedback on. This feedback is essential for planning process of the upcoming iterations as its shapes the scope and direction of the software development effort.
- **Human centric** [7]**:** meaning the reliance on people and the interaction between them is a cornerstone in the definition of agile software processes.
- **Technical excellence** [12, 10]**:** is necessary to ensure high quality. Agile team members are committed to producing only the highest quality code they can.  Maintaining high-quality is essential when developing at high speed environments as that of agile development processes.
- **Customer collaboration** [5]**:** inspired from the original statement of the agile manifesto, there must be significant and frequent interaction between the customers, developers, and all the stakeholders of the project to ensure that the product being developed satisfies the business needs of the customer.

## *Agile Practices and Concepts*

Agile Practices and Concepts are specific and practical techniques and steps by which the agile principles are realized.

For example: *paired programming* is an agile practice, *user stories* is another agile practice, *collaborative planning* is an agile concept.

***Within this agile Measurement Index, agile practices and concepts are used to characterize and exemplify the degree of adherence of an agile principle within an agile level***

**For example:**

Agile level 1 focuses on enhancing communication and cooperation. This agile value (level) is realized by adhering to various agile principles, one of which is *Technical Excellence.* Technical excellence is demonstrated in Agile Level 1 through a set of practices and concepts that exemplify basic technical excellence with a focus on communication and collaboration. For example these agile practices and concepts that are related to technical excellence and focus on communication and collaboration are:

- coding standards
- usage of knowledge sharing tools
- task Volunteering instead of task assignment

While at agile level 4 technical excellence will be exemplified through a different set of agile practices and concepts that promote responding to change through multiple levels of feedback, such as:

- daily progress tracking meetings
- agile documentation
- user stories

Each level of agility is satisfied when an agile value is added to the software development process. These level are achieved through the adherence to agile principles. The degree and scope of adherence varies depending on the level and are exemplified through different agile practices and concepts. Moving to the next level of agility implies that the agile practices and concepts in the current level of agility were all successfully adopted.

Table 5 below illustrates each of the Agile levels populated with practices and concepts categorized under their agile principles.

| | Agile Principles | | | | |
|---|---|---|---|---|---|
| | Delivering Customer Value by Embracing Change [10, 4] | Planning and Delivering Software Frequently [5] [16, 9] | Human Centric [7] | Technical Excellence [12, 10] | Customer Collaboration [5] |
| Level 1: **Collaborative [7, 8]**<br><br>*Theme: Communication* | Reflect and tune Process [14, 18] | Collaborative Planning [16, 8, 13] *(Customers, Developers and Management plan together)* | Collaborative teams<br><br>Empowered and Motivated Teams [5] | Coding Standards [11, 19, 15]<br><br>Knowledge Sharing Tools [13] *(Wikis, Blogs)*<br><br>Task Volunteering not Task Assignment [13] | Customer Commitment to work with Developing Team [5] |
| Level 2: **Evolutionary [13]**<br><br>*Theme: Early and Continuous delivery* | Evolutionary Requirements [13] *(Not all requirements are elicited at the beginning of the project; they evolve and change)* | Continuous Delivery (Incremental-Iterative development) [13, 12, 10, 4] (*No Big Bang development, divide development into releases and each release into iterations*)<br><br>Planning at different levels [9, 12] *(Overall and detailed planning)* | | Software Configuration Management [12]<br><br>Tracking Iteration through Working Software [13]<br><br>No Big Design Up Front (BDUF) [1, 4] | Customer Contract reflective of Evolutionary Development [10, 14] |
| Level 3: **Effective [11, 8]**<br><br>*Theme: Producing Working Software* | Continuous Improvement [12, 4] | Risk Driven Iterations [13]<br><br>Maintain a list of all remaining features (Backlog) [12] | Self Organizing Teams [13, 16, 12, 8] *(Responsibilities are communicated to the whole team and the team divides the work and determines best way to accomplish)*<br><br>Frequent face-to-face communication between the team [16, 8, 5] | The use of True Object Oriented Design and Construction [12]<br><br>Continuous Integration [13]<br><br>Have around 30% of Cockburn Level 2 and Level 3 people on team [7, 6]<br><br>Automated Suite of Unit Tests | |
| Level 4 **Adaptive [10]**<br><br>*Theme: Responding to Change through multiple levels of feedback* | Client Driven Iterations [13] (*Client prioritizes the features and functionality to be developed in next iteration* )<br><br>Customer Satisfaction Feedback [14, 18] | Smaller and More Frequent Releases [14] (*4-8 Weeks*)<br><br>Adaptive Planning [13] *(Based on feedback, planning details of each iteration are developed immediately before the iteration - Don't stick to a plan but continuously plan)* | | Daily Progress Tracking Meetings [2]<br><br>Agile Documentation (from Agile Modeling) [17, 12]<br><br>User Stories | Collaborative, Representative, Authorized, Committed and Knowledgeable (CRACK) Customer Immediately Accessible [6]<br><br>Customer contract revolves around commitment of collaboration, not features [10, 14] |
| Level 5 **Ambient**<br><br>*Theme: Building Agile Environment & Surroundings* | Low Process Ceremony [13, 16] *(Process Ceremony is the level of paperwork involved with a process; e.g. change requests signed by 3 levels of management)* | Agile Project Estimation [9] | Ideal Agile Physical Setup [13] *(The team is in the same room, no cubicles)* | Test Driven Development<br><br>Paired Programming [2] [4]<br><br>No/minimal number of Cockburn Level -1 or 1b people on team [7, 6] | Frequent Face-to-face interaction between developers & Users [4](Collocated) |

**Table 5. The Matrix of the 5 Agile Levels**

# Appendix B:
# Indicator Aggregation

# *Indicator Aggregation*

The previous sections presented the three major stages of the process framework along with the Agile Levels. In each of the 3 stages of the process framework assessment was conducted using an assessment table similar to the one below.  This section focuses on presenting more detail on the evaluation methodology that is conducted after the results to the questions and observations in the assessment table are gathered.

| Category of Assessment | Area to be assessed | Characteristic(s) to be assessed | To determine: | Assessment Method | **Sample Indicators** |
|---|---|---|---|---|---|
| People | Developers | Interaction | Whether there exists any levels of interaction between people hence laying a foundation for more team work | Interviewing | I1, I2, I3 |
| | | Collectivism | Whether people believe in group work and helping others or are they just concerned about themselves | Interviewing | I4 |
| | | Buy-In | Whether people are willing to work in teams | Interviewing | I5, I6 |
| | | | Whether people recognize that their input is valuable in group work or not | Interviewing | I7, I8 |

This section presents the steps that will be conducted in order to evaluate the results gathered from the assessment table. These steps and this evaluation methodology is based off of the framework of the Evaluation Environment [3]. To use an automated tool to conduct evaluation step or to learn more about the evaluation environment please visit:  https://www.orcacomputer.com/ee.

**Step 1: Compute a weight for each indicator**

The first step is to assign a weight to each indicator. A weight is a fractional value between 0 and 1 that expresses the indicator's level of influence on the parent factor. The weights of all the indicators belonging to the same factor must sum to 1. We will assume that all the indicators have an equal weight, however evaluators are free to assign indicators higher weights than other.

Therefore looking at the first factor in the example above, the weights can be computed as follows (under the assumption all indicators have an equal influence of the parent factor)

**1** (sum of all weights) / **3** (number of indicators) = **0.334** (approximate weight per indicator)

**Step 2: Compute weighed interval**

After we computed the weight for each of the indicators, the next step is to compute the weighted intervals for each of the indicators. For the above example we will assume the following answers were given to the sample indicators of the first factor being assessed.

| | | Normalized Categories | | | | |
|---|---|---|---|---|---|---|
| | | V | W | X | Y | Z |
| Indicator Number | Sample Question | **0%-15%** | **15%-40%** | **40%-60%** | **60%-85%** | **85%-100%** |
| I1 | Question 1 | X | | | | |
| I2 | Question 2 | | | X | | |
| I3 | Question 3 | | | | | X |

X Represents the answer that was chosen for that indicator

Once you have the answers from the sample indicators the next step is to multiple the weight of the indicator by the high and low end of the interval range selected for the indicator

| Indicator Number | Computed Weight | Interval Low End | Interval High End | Interval Low End X Weight | Interval High End X Weight |
|---|---|---|---|---|---|
| I1 | 0.33334 | 0 | 15 | 0 X 0.3334 = **0** | 15 X 0.3334 = **5** |
| I2 | 0.33334 | 40 | 60 | 40 X 0.3334 = **13** | 60 X 0.3334 = **20** |
| I3 | 0.33334 | 85 | 100 | 85 X 0.3334 = **28** | 100 X 0.3334 = **33** |

---

**Step 3: Calculate Result Range**

The next step is to compute the Result Range by calculating the optimistic and pessimistic range for each factor. This is accomplished by summing up all the weighed intervals we got from the previous step. The example below highlights in more detail how this is done.

Pessimistic Result = Sum of all the weighed low end results from Step 2
Pessimistic Result: 0 + 17 +28 = 41
Optimistic Result = Sum of all the weighed high end results from Step 2
Optimistic Result: 5 +28 + 33 = 58

**Your Result Range = 41 – 58**

---

**Step 4: Translate to Nominal Score**

In many cases the assessment table indicates that several aspects need to be assessed in order to completely assess a certain characteristic of a factor. In those cases we do not compute and nominal value, rather we perform another round of aggregation, as demonstrated above, but on the next level up till we reach the level of the characteristic being assessed. In the assessment table presented earlier you can see that to assess Interaction and Collectivism we do not have to go through another cycle of aggregation. However to assess the buy-in we have to aggregate the indicators and then we have to aggregate the 2 different aspects of buy-in that we are assessing before we can move to the step that determines the nominal assessment result for the characteristic being assessed.

Once you have a result range for that a particular characteristic, and you are sure you do not have to perform more aggregation the next step is to map the result range to one of the nominal values presented below. These nominal values are the ones that are used to evaluate the fulfillment of a particular factor or not.

| | |
|---|---|
| **Not Achieved** | **0%-35%** |
| **Partially Achieved** | **35%-65%** |
| **Largely Achieved** | **65%-85%** |
| **Fully Achieved** | **85% - 100%** |

**Table 6. Nominal Values**

If the Pessimistic - Optimistic (From Step 3) range fits within one of these intervals then that suffices, if they do not then obtain an average and then place that average in its necessary nominal range.
In our example the resultant score will be: **Partially Achieved**
Below is a sample of the evaluation template that would be used for the assessment table example given earlier

| Category of Assessment | Area to be assessed | Characteristic(s) to be assessed | **Nominal Value** | Weight | Low | High | Indicator | Weight | Low | High |
|---|---|---|---|---|---|---|---|---|---|---|
| Project Management | Developers | Interaction | | 1 | | | I1 | 0.333 | | |
| | | | | | | | I2 | 0.333 | | |
| | | | | | | | I3 | 0.333 | | |
| | | Collectivism | | 1 | | | I4 | 1.000 | | |
| | | Buy-In | | 0.5 | | | I5 | 0.500 | | |
| | | | | | | | I6 | 0.500 | | |
| | | | | 0.5 | | | I7 | 0.500 | | |
| | | | | | | | I8 | 0.500 | | |

**Table 7. Evaluation Template for an Assessment Table**

Once a nominal score is reached for each characteristics being assessed, their nominal values are plugged in to the evaluation matrix similar to the Table 8 below to determine which areas need to be addressed before trying to adopt that particular agile practice.

| Agile Practices for Agile Level 1 | Category of Assessment | Area to be assessed | Characteristic(s) to be assessed | Not Achieved | Partially Achieved | Largely Achieved | Fully Achieved |
|---|---|---|---|---|---|---|---|
| | | | | 0%-35% | 35%-65% | 65%-85% | 85% - 100% |
| | | | | Large Gap | Medium Gap | Small Gap | Minimal Gap |
| **Collaborative Planning** | People | Management | Management Style | | X | | |
| | | | Buy-In | X | | | |
| | | | Transparency | | | X | |
| | | Developers | Power Distance | | | | X |
| | | | Buy-In | | | | X |
| | Project Management | Planning | Existence | | | X | |
| **Task Volunteering not Task Assignment** | People | Management | Buy-In | | | | X |
| | | Developers | Buy-In | | | X | |
| **Collaborative Teams** | Project Management | Developers | Interaction | | X | | |
| | | | Collectivism | | | X | |
| | | | Buy-In | | | | X |
| **Empowered and Motivated Teams** | People | Developers | Decision Making | | X | | |
| | | | Motivation | | | X | |
| | | Managers | Trust | | | | X |
| **Coding Standards** | People | Developers | Buy-In | | | | X |
| | Process | Coding Standards | Existence | | | | X |
| **Knowledge Sharing** | People | Developers | Buy-In | | | X | |
| | | Managers | Buy-In | | | X | |
| | Process | Knowledge Sharing | Availability | X | | | |

**Table 8. Sample Evaluation Matrix for Agile Level 1**